\documentclass[twocolumn]{aastex631}

\usepackage{multirow}

\usepackage{color}

\usepackage[normalem]{ulem}

\usepackage[utf8]{inputenc}
\usepackage{natbib}
\usepackage{graphicx}
\usepackage{longtable}
\usepackage{enumerate}
\usepackage{amsmath}
\usepackage{hyperref}
\usepackage{CJK}
\usepackage[title]{appendix}
\usepackage{rotating}
\usepackage{longtable}

\def\msun{\ifmmode {\rm M}_{\mathord\odot}\else $M_{\mathord\odot}$\fi}
\def\rsun{\ifmmode {\rm R}_{\mathord\odot}\else $R_{\mathord\odot}$\fi}
\def\lsun{\ifmmode {\rm L}_{\mathord\odot}\else $L_{\mathord\odot}$\fi}

\def\co{$^{12}$CO}
\def\c18o{C$^{18}$O}
\def\h2{H$_{2}$}
\def\13co{$^{13}$CO}
\def\n2hp{$_{2}$H$^{+}$}

\def\cm2{cm$^{-2}$}

\newcommand{\um}{$\mu$m}

\def\um{$\mu$m}

\newcommand{\CASItD}{{\sc casi-3d}}

\shorttitle{}
\shortauthors{}

\begin{document}
\begin{CJK*}{UTF8}{gbsn}

\title{A Census of Outflow to Magnetic Field Orientations in Nearby Molecular Clouds}

\author[0000-0001-6216-8931]{Duo Xu}
\affiliation{Department of Astronomy, University of Virginia, Charlottesville, VA 22904-4235, USA}

\author[0000-0003-1252-9916]{Stella S. R. Offner}
\affiliation{Department of Astronomy, The University of Texas at Austin, Austin, TX 78712, USA}

\author[0000-0002-6447-899X]{Robert Gutermuth}
\affiliation{Department of Astronomy, University of Massachusetts, Amherst, MA 01003, USA }

\author[0000-0002-3389-9142]{Jonathan C. Tan}
\affiliation{Department of Astronomy, University of Virginia, Charlottesville, VA 22904-4235, USA}
\affiliation{Department of Space, Earth \& Environment, Chalmers University of Technology, SE-412 96 Gothenburg, Sweden}


\email{xuduo117@virginia.edu}

\begin{abstract}
We 
define a sample of 200 protostellar outflows showing blue and redshifted CO emission in the nearby molecular clouds Ophiuchus, Taurus, Perseus and Orion to investigate the correlation between outflow orientations and local, but relatively large-scale, magnetic field directions traced by {\it Planck} 353 GHz dust polarization. At high significance ($p\sim 10^{-4}$), we exclude a random distribution of relative orientations and find that there is a 
preference for alignment of projected plane of sky outflow axes with magnetic field directions. The distribution of relative position angles peaks at $\sim 30^{\circ}$ and exhibits a broad dispersion of $\sim 50^\circ$. These results indicate that magnetic fields have dynamical influence in regulating the launching and/or propagation directions of outflows. However, the significant dispersion around perfect alignment orientation implies that there are large measurement uncertainties and/or a high degree of intrinsic variation caused by other physical processes, such as turbulence or strong stellar dynamical interactions.
Outflow to magnetic field alignment is expected to lead to a correlation in the directions of nearby outflow pairs, depending on the degree of order of the field. Analyzing this effect we find limited correlation, except on relatively small scales $\lesssim 0.5\:$pc. Furthermore, we train a convolutional neural network to infer the inclination angle of outflows with respect to the line of sight and apply it to our outflow sample to estimate their full 3D orientations. We find that the angles between outflow pairs in 3D space also show evidence of small-scale alignment. 

\end{abstract}


\section{Introduction}
\label{Introduction}

Protostellar outflows play a crucial role in star formation. They inject a substantial amount of mass, momentum and energy into the surroundings, which heat and compress the ambient gas and may offset the rapid turbulent dissipation of the host molecular cloud \citep[e.g.,][]{2016MNRAS.457..375F,2016ARA&A..54..491B}. Moreover, protostellar outflows significantly reduce protostellar masses and accretion rates, substantially affecting the shape of the stellar initial mass function in magneto-hydrodynamic simulations \citep[IMF,][]{2015MNRAS.450.4035F,2017ApJ...847..104O,2018MNRAS.476..771C,2021MNRAS.502.3646G}.

Protostellar outflows are lauched by the interplay of a rotating accretion disk and magnetic fields \citep[e.g.,][]{2016ARA&A..54..491B}. Both numerical and observational studies suggest that outflows are launched parallel to the angular momentum vector of the accretion disk \citep{2002ApJ...575..306T,2004ApJ...616..266M,2009A&A...494..147L}. However,  the role magnetic fields play in setting the outflow orientation is still under debate. \citet{2004ApJ...616..266M} used ideal magneto-hydrodynamic simulations to investigate how different magnetic strengths affect the alignment between the outflow orientation and magnetic field direction. They concluded that when magnetic fields are strong ($B\ge40\, \mu$G), the outflow is well aligned ($\sim 5^{\circ}$) with the magnetic field of the parent cloud, while in a weak magnetic field scenario ($B\le20\, \mu$G), the outflow is still aligned with the magnetic field on average but with larger scatter ($\sim 30^{\circ}$).  \citet{2017ApJ...834..201L} carried out a similar numerical study including magnetized turbulence. They found that both a weaker field and dynamical interactions reduce the correlation between the outflow and field directions, resulting in a near random distribution of angles. In contrast, \citet{2020MNRAS.491.2180M} conducted resistive magnetohydrodynamics simulations to study the launching of protostellar outflows and found no correlation between outflow orientation and magnetic field direction, even in the absence of any turbulence during the early phase of star formation. 

The observational evidence is similarly ambiguous. The orientation between outflows and magnetic fields appears random in NGC 1333 \citep{2020ApJ...899...28D}, but not in IRDC G28.37+0.07, where most outflows have a preferential orientation that is consistent with the direction of the magnetic field \citep{2019ApJ...874..104K}. Moreover, \citet{2021ApJ...907...33Y} found that the relative position angle between outflows and magnetic fields has a typical value of 15$^{\circ}$-35$^{\circ}$, indicating a moderate degree of alignment.

Several factors might explain these inconsistent findings. First, the small scale magnetic field direction is not easy to measure \citep{2019FrASS...6....3H}. In addition, different studies measure the field on different scales. The outflow orientation is also likely affected by dynamical interactions in denser star-forming regions such as NGC 1333, where the signature of any outflow-magnetic field correlation may then be erased \citep{2016ApJ...827L..11O,2017ApJ...834..201L}. Moreover, the magnetic field strength varies between clouds, so that one cloud with strong fields may appear to have more aligned outflows while another with weaker fields may have more randomly oriented outflows \citep{2004ApJ...616..266M,2017ApJ...834..201L}.

In this work, we aim to investigate some of these factors by carrying out a large statistical sample including outflows in a range of different environments. We adopt a set of outflows identified by a supervised machine learning approach \citep{2019ApJ...880...83V,2020ApJ...905..172X}. 
\citet{2020ApJ...905..172X} developed a Convolutional Approach to Shell/Structure Identification-3D, \CASItD, to identify protostellar outflows in position-position-velocity (PPV) molecular line spectral cubes. \citet{2022ApJ...926...19X} applied \CASItD\ 
to create a census of protostellar outflows in the nearby molecular clouds, Ophiuchus, Taurus, Perseus and Orion. These clouds span a range of different column densities, gas properties and stellar densities. 
 
In this paper, we adopt the highest confidence outflow candidates identified by \citet{2022ApJ...926...19X} in these four nearby clouds to study the correlation between outflows and magnetic fields. We describe the outflow sample and Planck dust polarization data in Section~\ref{Data and Method}. We introduce a new \CASItD\ model that we train to infer the inclination angles of the outflows with respect to the line of sight in Section~\ref{Data and Method}. In Section~\ref{Results}, we present a statistical study of the outflow orientations, and we summarize our results and conclusions in Section~\ref{Conclusions}.

\section{Data and Method}
\label{Data and Method}

\subsection{Outflow Candidates}
\label{Outflow Candidates}

\begin{figure*}[hbt!]
\centering
\includegraphics[width=0.78\linewidth]{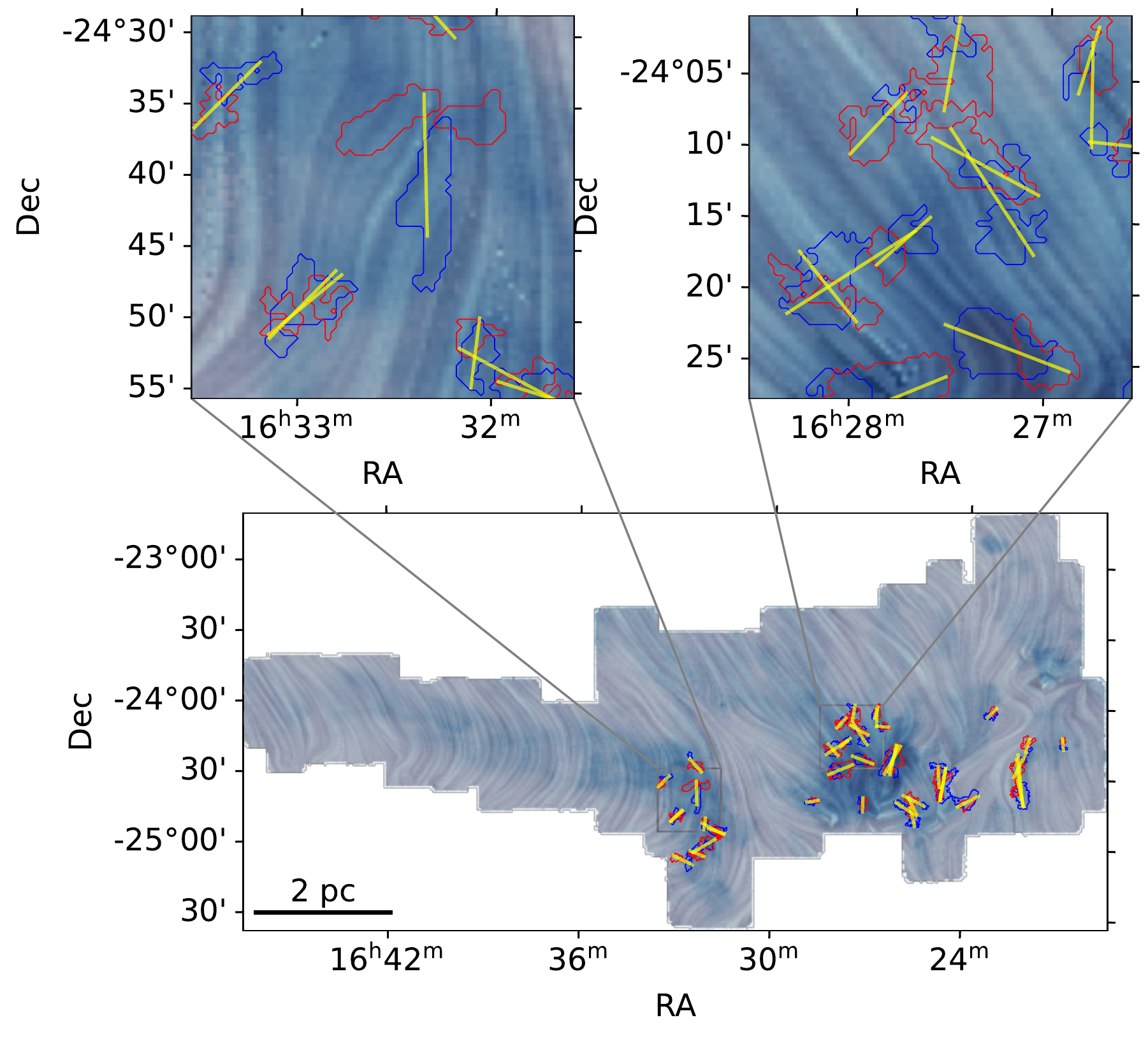}
\caption{\co\ $(1-0)$ integrated intensity of Ophiuchus overlaid with the magnetic field streamlines derived from the Planck dust emission. Blue and red contours indicate the outflow lobes. Yellow lines indicate the orientation of the outflow lobe pairs.}
\label{fig.hist-all-outflow-bfield-oph}
\end{figure*}

In this analysis, we consider the protostellar outflows identified by \CASItD\ in the four molecular clouds, Ophiuchus, Taurus, Perseus and Orion. \citet{2022ApJ...926...19X} employed \CASItD\ to systematically identify protostellar outflows in \co\ $(1-0)$ and \13co\ $(1-0)$ data cubes. The \co\ and \13co\ data of Ophiuchus, Taurus and Perseus were observed with the 13.7 m Five College Radio Astronomy Observatory (FCRAO) Telescope \citep{2006AJ....131.2921R, 2008ApJS..177..341N}. The main beam of the antenna pattern has a FWHM of 45\arcsec\ for \co\ and 47\arcsec\ for \13co. The data are obtained on the fly (OTF), but they are resampled onto a uniform 23\arcsec\ grid \citep{2006AJ....131.2921R}. The observations of Orion A were carried out with the Nobeyama Radio Observatory 45 m telescope (NRO 45m) \citep{2015ApJS..217....7S,2015ApJS..221...31S,2019PASJ...71S...9I,2019PASJ...71S...3N}. The data has a pixel scale of 7\arcsec.5 and has an effective angular resolution of 22\arcsec. {There is a span of distance estimates for the four clouds. We adopt fiducial distance estimates for the four clouds of 120 pc for Ophiuchus \citep{2011ApJ...726...46N}, 140 pc for Taurus \citep{2012MNRAS.425.2641N}, 300 for Perseus \citep{2010ApJ...715.1170A} and 420 for Orion \citep{2018ApJS..236...25K}. The physical resolutions for the four clouds are 0.013 pc per pixel for Ophiuchus, 0.016 pc per pixel for Taurus, 0.033 pc per pixel for Perseus and 0.015 pc per pixel for Orion.} 

{
\CASItD\ is an encoder-decoder based 3D convolutional neural network, which identifies outflow structures coherently across velocity channels. This indicates that \CASItD\ identifies outflows using both morphology and velocity information from molecular line data cubes. We train \CASItD\ on the same training set as that in \citep{2020ApJ...905..172X}, which includes different magnetohydrodynamic model properties, different \co\ abundances, and different cloud kinetic temperatures. \CASItD\ takes \co\ data cubes as input and predicts the
position of outflows on the voxel level.} We separate the outflow prediction into two components: blue-shifted lobes and red-shifted lobes for each cloud. We exclude the emission near the central velocity where $|v_{\rm cen}|<1$ km/s. We adopt the integrated blue- and red-shifted outflow components predicted by model MF, a model that is trained to exclude the contamination by emission that is not associated with feedback \citep[e.g.,][]{2020ApJ...905..172X,2020ApJ...890...64X}. 

\CASItD\ identifies all the voxels associated with feedback, but does not segment them into individual outflows. 
{Therefore, after applying \CASItD}, we carry out a dendrogram analysis{\footnote{https://dendrograms.readthedocs.io/en/stable/}} on the outflow prediction to isolate individual outflow lobes. We vary several different parameters to verify {that} the results do not strongly depend on the assumed values {used to construct the dendrogram tree}. For example, in {analyzing} Perseus, we vary the {\it min\_value} parameter between 7 and 10~$\sigma$, {\it min\_delta} between 1 and 2~$\sigma$, and {\it min\_npix} between 10 and 20 pixels. {The parameter {\it min\_value} is the minimum value to consider in constructing the tree, and our values are similar to the detection level of the observations. The parameter {\it min\_delta} indicates how significant a leaf has to be in order to be considered an independent entity. For observational data this is usually set to 1~$\sigma$, which means that any leaf that is locally less than 1~$\sigma$ tall is combined with its neighboring leaf or branch and is no longer considered a separate entity. The parameter {\it min\_npix} is the minimum number of pixels needed for a leaf to be considered an independent entity. The leaf will be joined to its parent branch or another leaf if the leaf has fewer than this number of pixels.} The number of identified blue-shifted lobes ranges between 102 and 111, and the number of identified red-shifted lobes is between 58 and 62. The relatively bright lobes are universally identified regardless of the tested parameter. 

\begin{figure*}[hbt!]
\centering
\includegraphics[width=0.78\linewidth]{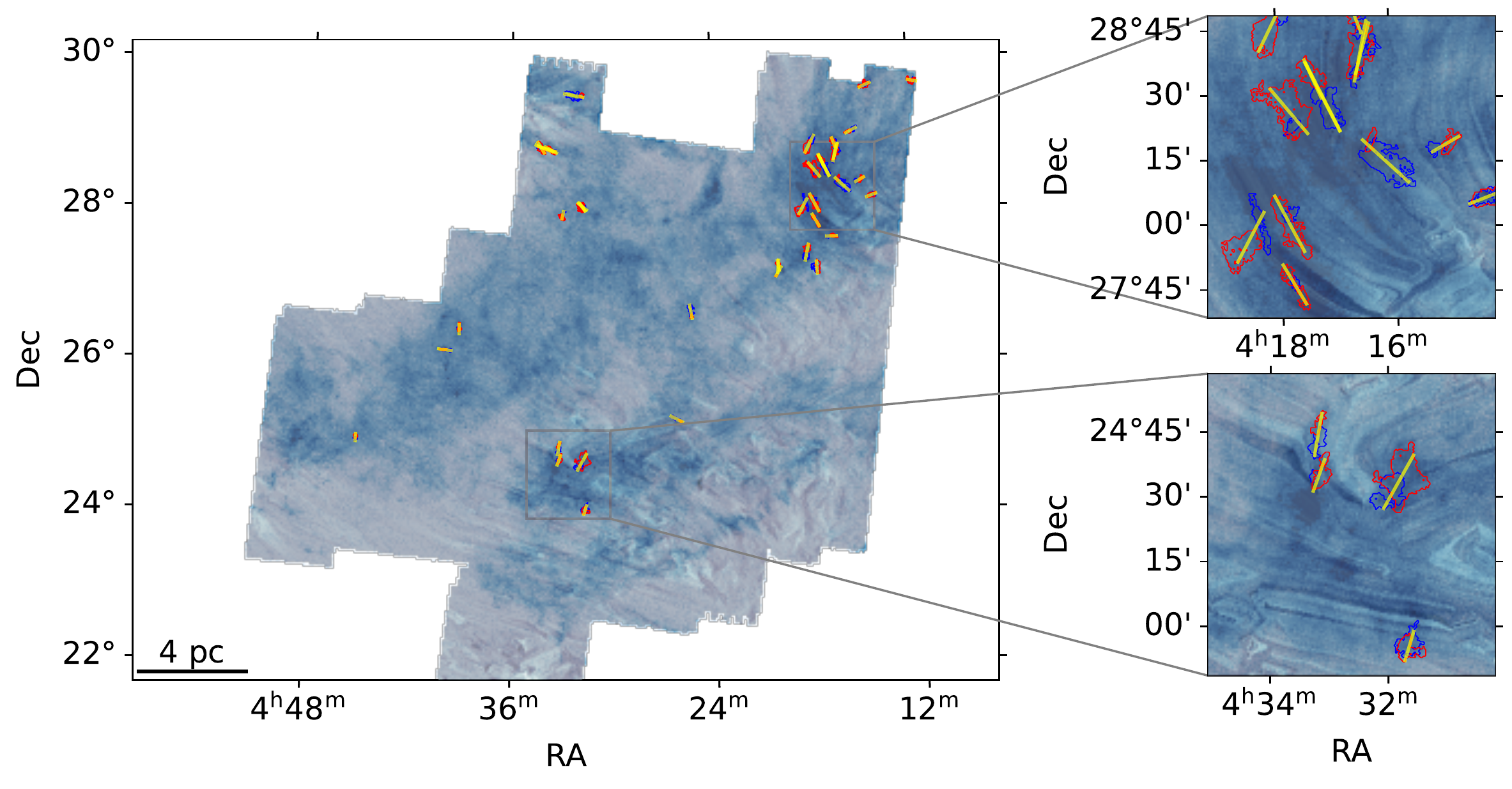}
\caption{Same as Figure~\ref{fig.hist-all-outflow-bfield-oph}, but for Taurus.}
\label{fig.hist-all-outflow-bfield-taurus}
\end{figure*}

\begin{figure*}[hbt!]
\centering
\includegraphics[width=0.78\linewidth]{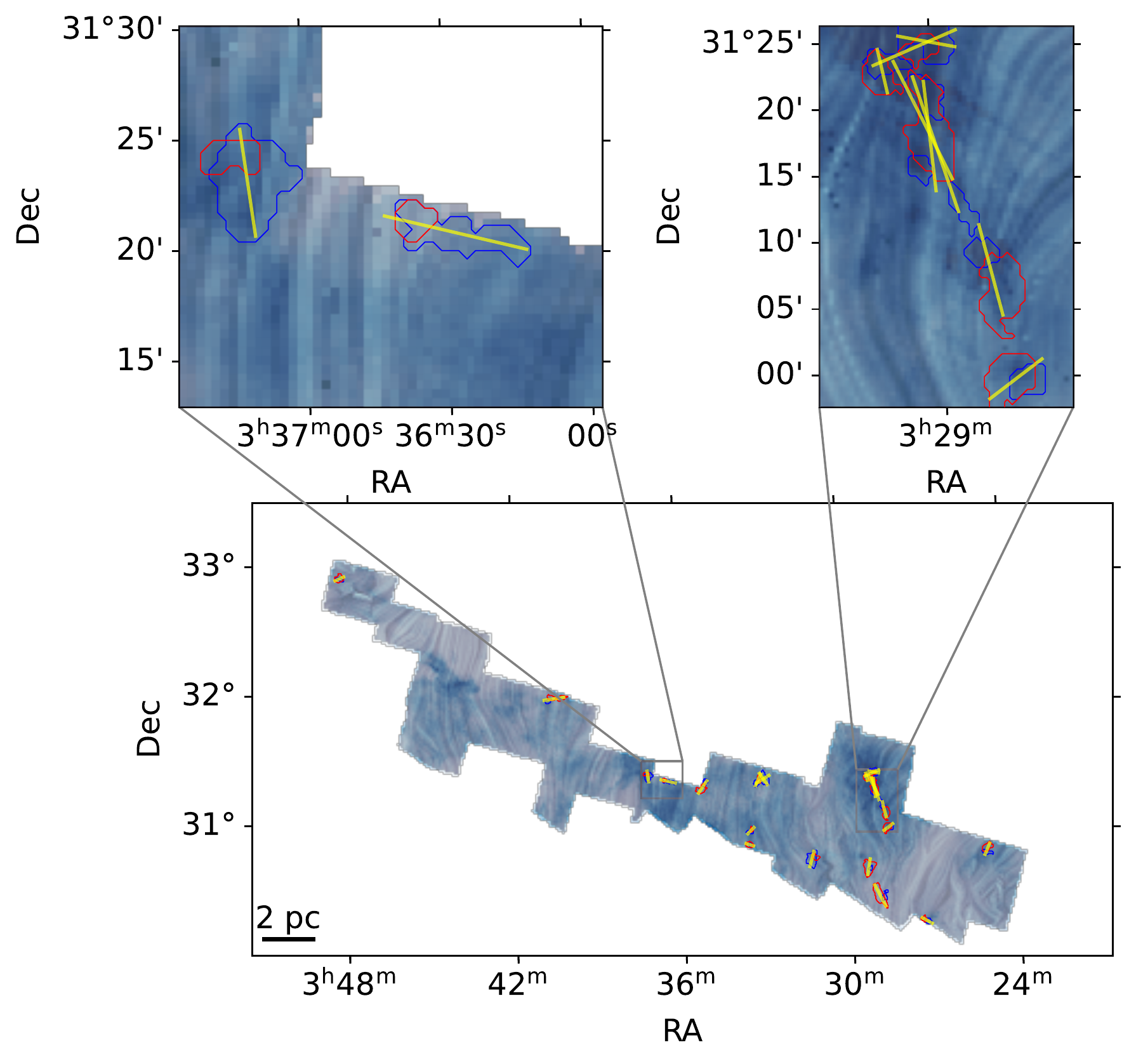}
\caption{Same as Figure~\ref{fig.hist-all-outflow-bfield-oph}, but for Perseus.}
\label{fig.hist-all-outflow-bfield-perseus}
\end{figure*}

\begin{figure}[hbt!]
\centering
\includegraphics[width=0.98\linewidth]{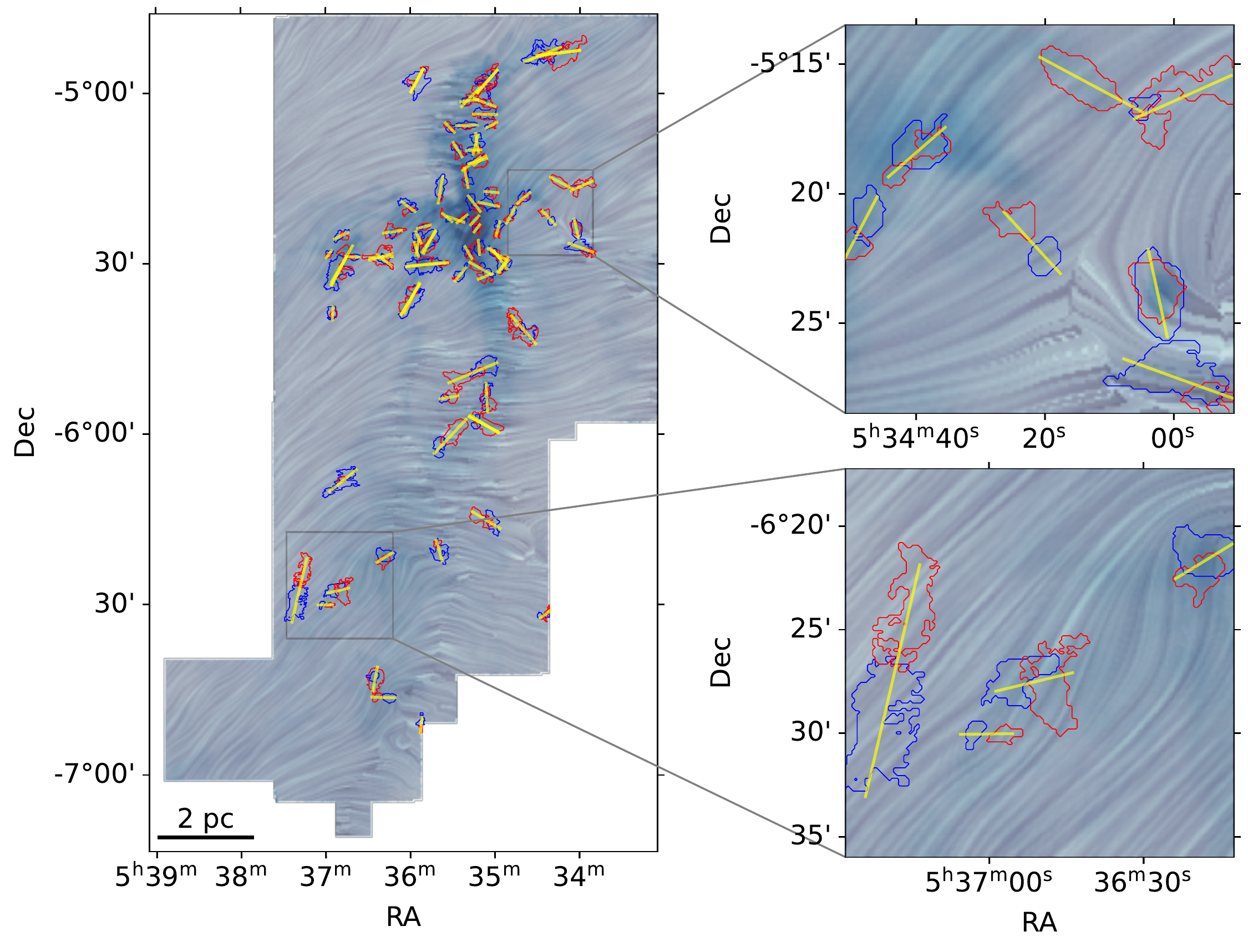}
\caption{Same as Figure~\ref{fig.hist-all-outflow-bfield-oph}, but for Orion. }
\label{fig.hist-all-outflow-bfield-orion}
\end{figure}

Many of the outflow features are identified in clustered regions where it is difficult to define the outflow direction. To address this, we pair blue- and red-shifted lobes if their distance is within 0.1 pc and then derive a single outflow orientation using both lobes. We limit our analysis to this set of candidates and exclude all one-lobe outflows, which only have either a blue- or red-shifted lobe. We note that it is common for only one lobe of the outflow to appear cleanly in CO data  \citep[e.g.,][]{2010ApJ...715.1170A}, however at the data resolution one outflow lobe alone may not have a clear orientation. The outflows tend to be more isolated. We apply principal component analysis (PCA) to the outflow pairs to determine the orientations. Figures~\ref{fig.hist-all-outflow-bfield-oph}-\ref{fig.hist-all-outflow-bfield-orion} show the outflow pairs in the four regions. Hereafter, we use the term outflow to refer to identified pairs of blue- and red-shifted lobes. Altogether, we identify 43 outflows in Ophiuchus, 41 in Taurus, 23 in Perseus and 93 in Orion, for a total of 200 outflows in the four regions. We further examine the position-velocity diagram of all 200 outflow candidates. Of these, 136 outflow candidates have significant coherent high velocity features across at least 1 km/s, with a characteristic ``Hubble wedge" shape in the position-velocity diagram \citep[][see supplementary images]{2001ApJ...551L.171A}. We consider these 136 outflow candidates as a subset of the highest-confidence outflows. We retain the rest as likely outflow candidates, which have distinct blue- and redshifted lobes but do not show significantly coherent velocity structure across at least 1 km/s. However, these outflow candidates are likely real outflows, since many enclose an infrared source. \citet{2015ApJS..219...20L} found a significant number of outflows in Taurus that do not have over 1 km/s coherent velocity structure in the position-velocity diagram, but they are clear across several channels. Consequently, we adopt all 200 outflow candidates that have both blue- and redshifted lobes as the primary sample in our following work.

\subsection{Planck 353 GH Dust Polarization Map}
\label{Planck 353 GH Dust Polarization Map}

We adopt the data from the Planck 3rd Public Data Release \citep{2020A&A...641A..12P}. We infer the magnetic field orientation from the dust polarization angle:
\begin{align}
\label{phi-B}
\phi_{B} =\frac{1}{2} {\rm arctan2} (-U,Q)+\frac{\pi}{2},
\end{align}
where Q and U are the Stokes parameters of polarized dust emission.
The maps of Q and U are initially at 4$^{\prime}$.8 resolution in HEALPix
format with an effective pixel size of 1$^{\prime}$.07. We calculate the magnetic field orientation of each outflow by taking the mass-weighted magnetic field orientation at the corresponding position. We show the magnetic field orientation of the four clouds in Figures~\ref{fig.hist-all-outflow-bfield-oph}-\ref{fig.hist-all-outflow-bfield-orion}. 

\subsection{\CASItD: Inferring Outflow Inclination Angles }
\label{sec-CASItD}

\begin{figure}[hbt!]
\centering
\includegraphics[width=0.98\linewidth]{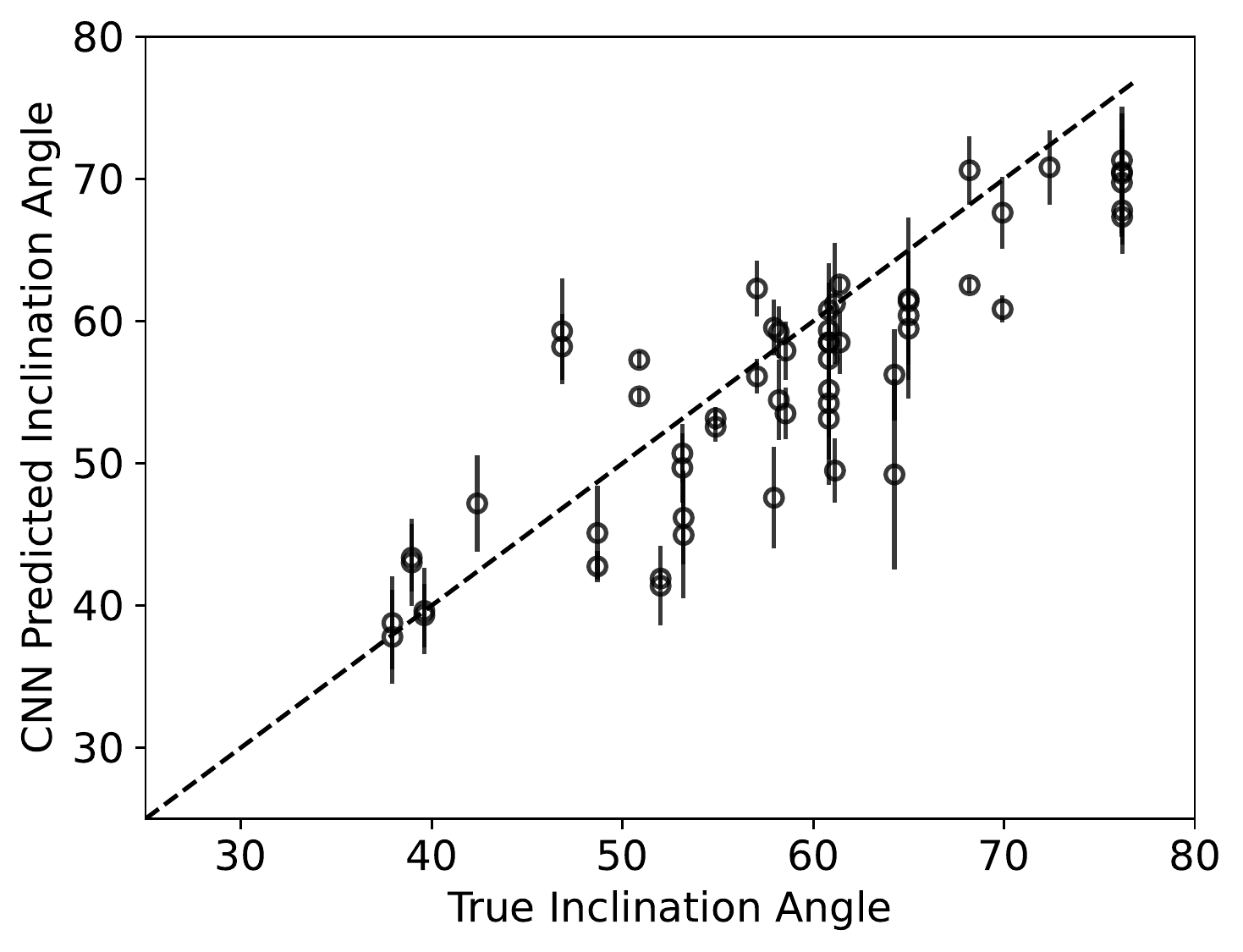}
\caption{Relation between the \CASItD\ predicted outflow inclination angles and the
true inclination angles for synthetic outflows. The error bars indicate the uncertainty of the inferred inclination angle for synthetic outflows with different physical and chemical conditions.}
\label{fig.test-inclination-errorbar}
\end{figure}

\begin{figure*}[hbt!]
\centering
 \includegraphics[width=0.48\linewidth]{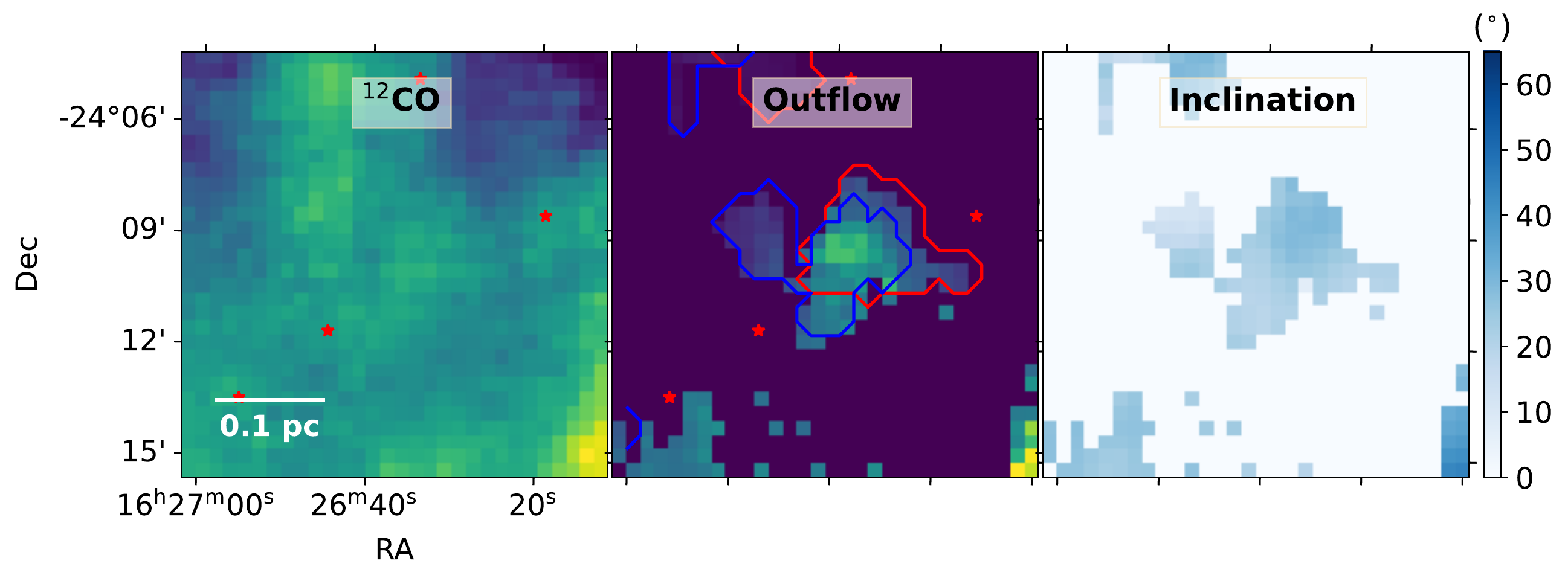}
\includegraphics[width=0.48\linewidth]{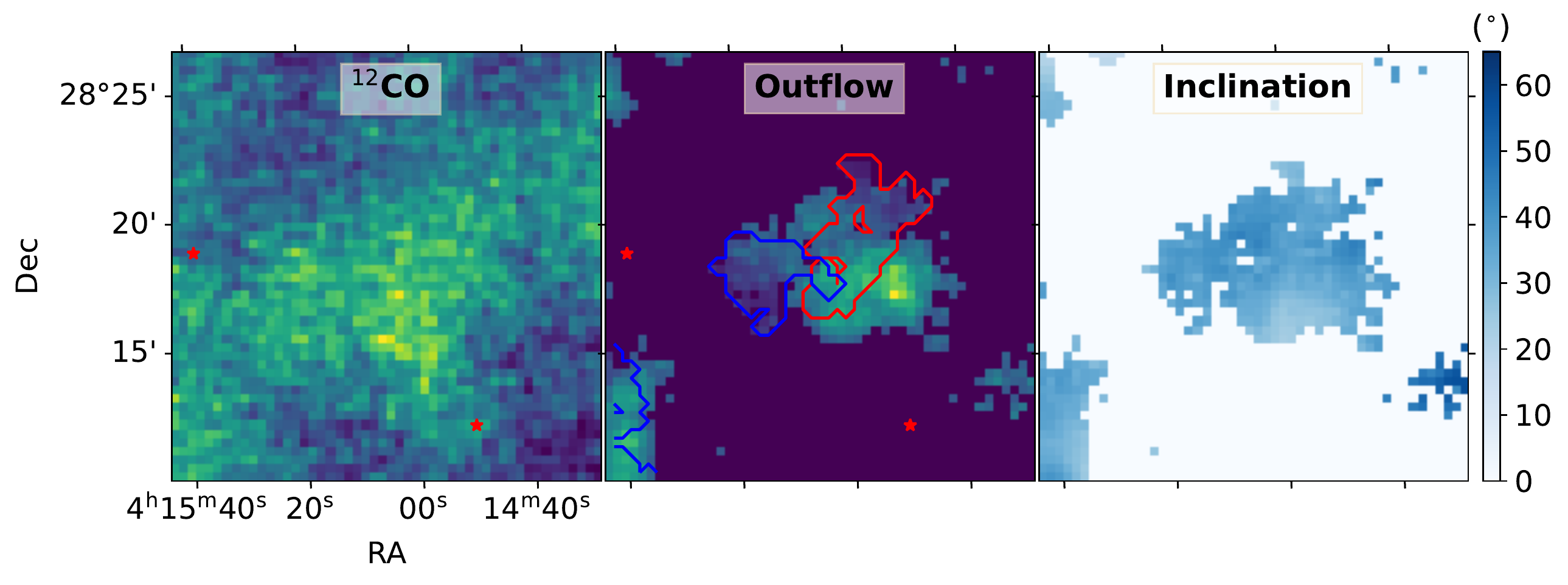}
\includegraphics[width=0.48\linewidth]{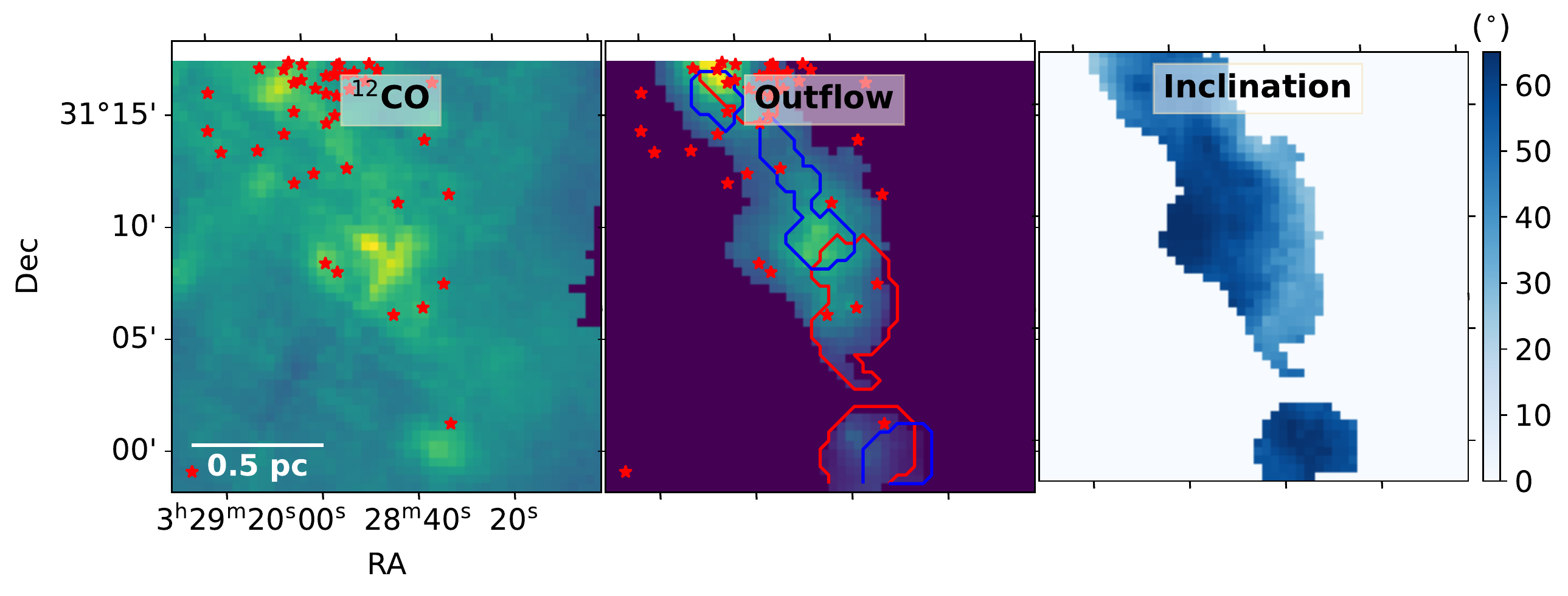}
 \includegraphics[width=0.48\linewidth]{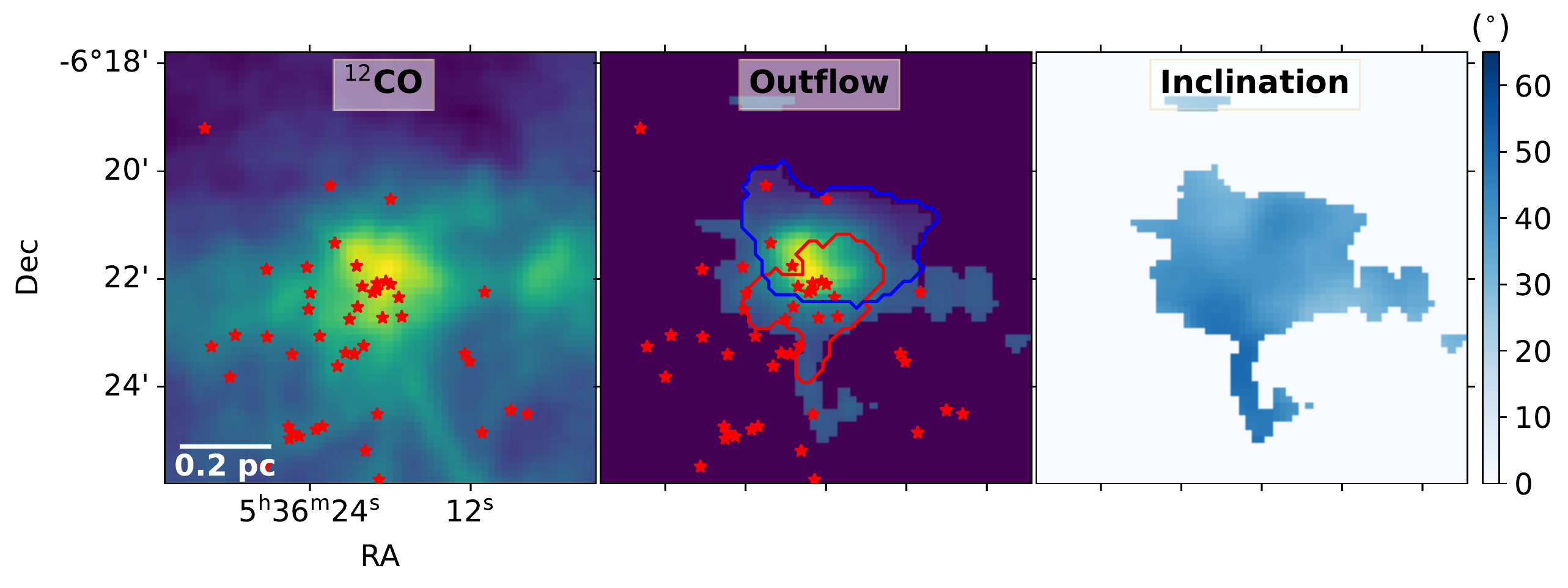}

\caption{\CASItD\ prediction for previously identified outflows in Ophiuchus (upper left), Taurus (upper right), Perseus (lower left) and Orion (lower right). Left: integrated \co\ emission. Middle: outflow emission predicted by \CASItD. Blue and red contours indicate the blue- and red-shifted lobes. Right: inclination angle predicted by \CASItD\ across the outflows. The effective outflow inclination angles are 23$^{\circ}$ (upper left), 36$^{\circ}$ (upper right), 48$^{\circ}$ (lower left) and 40$^{\circ}$ (lower right). The red stars indicate the locations of YSOs \citep{2009ApJS..184...18G,2010ApJS..186..259R,2020ApJ...896...60P}. }
\label{fig.outflow-per-tau-outflow-example-casi}
\end{figure*} 

\begin{figure}[hbt!]
\centering
\includegraphics[width=0.98\linewidth]{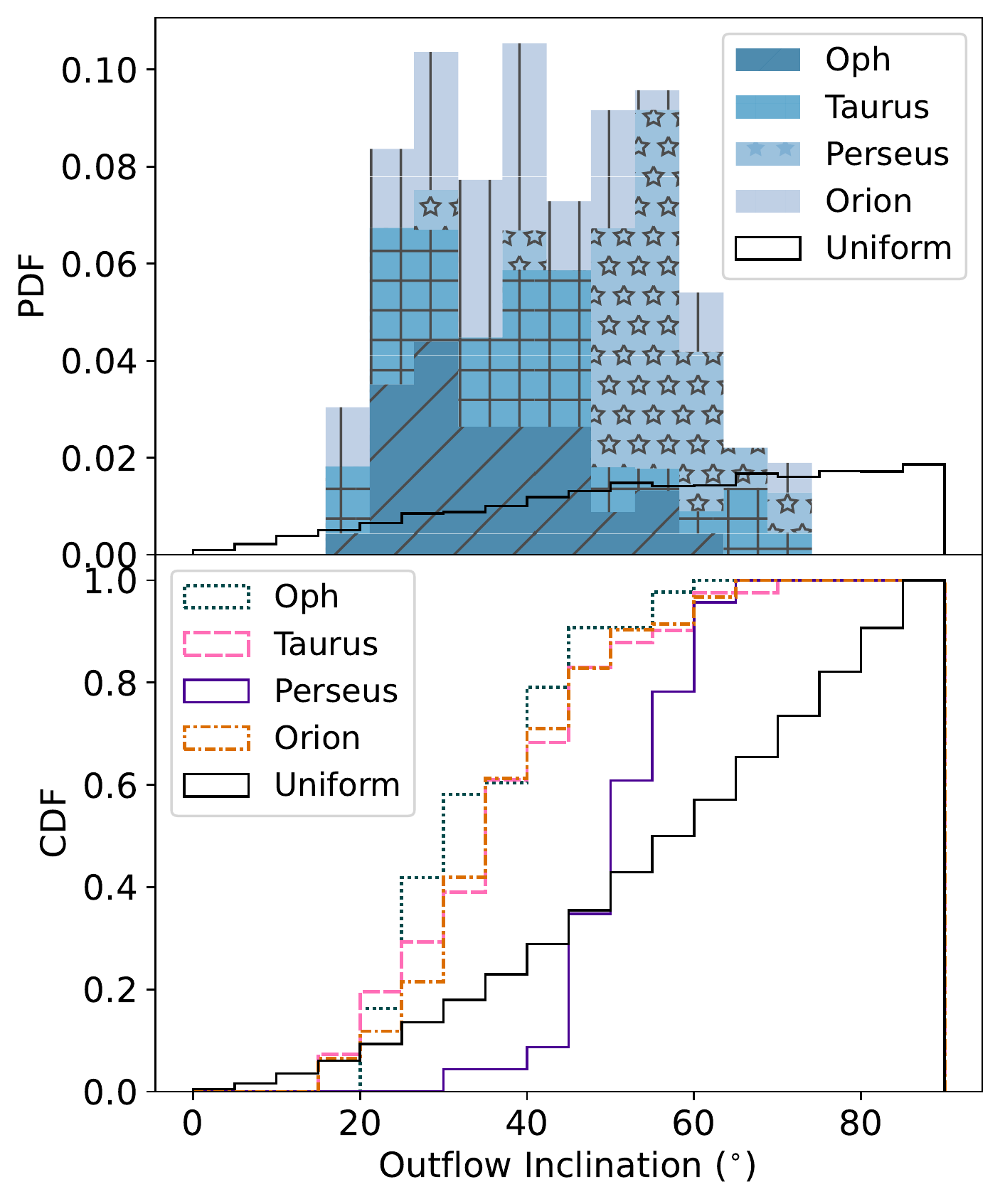}
\caption{{Stacked probability} distribution functions (PDF, top panel) and cumulative distribution functions (CDF, bottom panel) of the outflow inclination angles for the sources in the four regions.}
\label{fig.cdf-inclination-four-region}
\end{figure}

In this section, we introduce a new \CASItD\ model to predict the inclination angle of each protostellar outflow. 
We define $f_{v}$ as the ratio between the actual 3D momentum to the line-of-sight (LOS) momentum in each voxel,
 \begin{align}
\label{fv-eq}
f_{v} =\frac{P_{\rm 3D}}{P_{\rm LOS}}.
\end{align}
We adopt the same \CASItD\ architecture and training set as that in \citet{2020ApJ...905..172X}. We only replace the target from outflow mass to $f_{v}$ during training. The new model is able to predict the 3D momentum of the outflows in each voxel. We calculate both the 1D line-of-sight (LOS) and full 3D momentum of each outflow and define the effective outflow inclination angle, $i$, as
\begin{align}
\label{inclination-eq}
{\rm cos}\, i=\frac{\sum P_{\rm LOS}}{\sum P_{\rm 3D}}.
\end{align}
It is worth noting that this angle is not exactly equal to the geometric inclination angle of the outflow. Considering that most outflows are not pencil-beam jets in \co\ emission, the mass/momentum injection occurs over a wide-angle cone structure. This indicates that even if the outflow is launched perpendicularly to our LOS, we are still able to estimate the LOS momentum of the outflow cone, which implies ${\rm cos}\, i$ cannot be zero. Consequently, the effective inclination angle is an approximation of the geometric inclination angle of an outflow. 

Figure~\ref{fig.test-inclination-errorbar} demonstrates the performance of \CASItD\ for inferring the inclination angle of the synthetic outflows. The error bars indicate the uncertainty of the inferred inclination angle for synthetic outflows with different physical and chemical conditions, including different kinetic temperatures, 10 and 14 K, and different \co\ abundances, $10^{-4}$, $5 \times 10^{-5}$ and $10^{-5}$ \citep{2020ApJ...905..172X}. We discuss the performance of \CASItD\ in more detail in Appendix~\ref{CASI Performance}.

We next apply the \CASItD\ model to four high-confidence outflows 
{identified in Perseus, Taurus, Ophiuchus and Orion in order to show the performance for typical sources}. The inclination angle is calculated from the LOS momentum in each pixel and the corresponding 3D momentum predicted by \CASItD. For example, the inclination angle of the Ophiuchus outflow in the upper left panel of Figure~\ref{fig.outflow-per-tau-outflow-example-casi} is generally small compared with that of the other three outflows. This Ophiuchus outflow is likely launching towards us rather than in the plane of sky. The blue- and red-shifted lobes significantly overlap, which is consistent with an outflow that has a small inclination angle with respect to the line of sight. 



\section{Results}
\label{Results}

\subsection{Outflow Orientation Distributions}
\label{Outflow Orientation Distributions}

For each outflow identified in Section ~\ref{Outflow Candidates}, we adopt the total LOS momentum of the outflow and total \CASItD\ predicted 3D momentum to derive the inclination angle. We present the distribution of outflow inclination angles for the four regions in Figure~\ref{fig.cdf-inclination-four-region}. 

{With the exception of Perseus, the distributions are not consistent with a random distribution of inclinations. The outflows in Perseus have larger average inclination angles and a narrower distribution of angles than those in the other regions, i.e., they appear more randomly oriented. One possible explanation is that the plane-of-sky magnetic field might be weaker in Perseus than that in the other regions. When the magnetic field is strong, outflows likely launch along with the large-scale magnetic field direction rather than randomly. 
\citet{2020pase.conf..117B} summarized the recent results in the BISTRO (B-fields In STar-forming Regions Observations) polarimetric survey of several nearby molecular clouds with JCMT and found that the plane-of-sky magnetic field strength in Perseus B1 region is the weakest among the nearby molecular clouds. 
 However, B1 is a small subregion of Perseus and may not be representative of the conditions in the entire cloud. Another caveat is that these measurements are of the plane-of-sky magnetic field strength rather than the line-of-sight magnetic field strength. We note that the strength of the line-of-sight magnetic field at relatively large scales is poorly constrained for these four clouds.
 } 

Figure~\ref{fig.cdf-inclination-four-region} shows the average outflow inclination angle in Ophiuchus, Taurus and Orion is around $30^{\circ}-40^{\circ}$, which is smaller than the typical random value $60^{\circ}$. {Besides differences in the magnetic field or other cloud properties,} a possible explanation is that the \CASItD\ model has some bias towards identifying outflows with significant coherent high-velocity features. This implies that the identified outflow likely launches towards the line of sight rather than on the plane of sky, where the outflow gas blends with the cloud emission. Such an effect could introduce a bias that leads to a global signature of alignment in 3D, which we discuss further in the following sections.


\subsection{Outflow Orientation versus Magnetic Field Direction}
\label{Outflow Orientation versus Magnetic Field Directions}

\begin{figure*}[hbt!]
\centering
\includegraphics[width=0.455\linewidth]{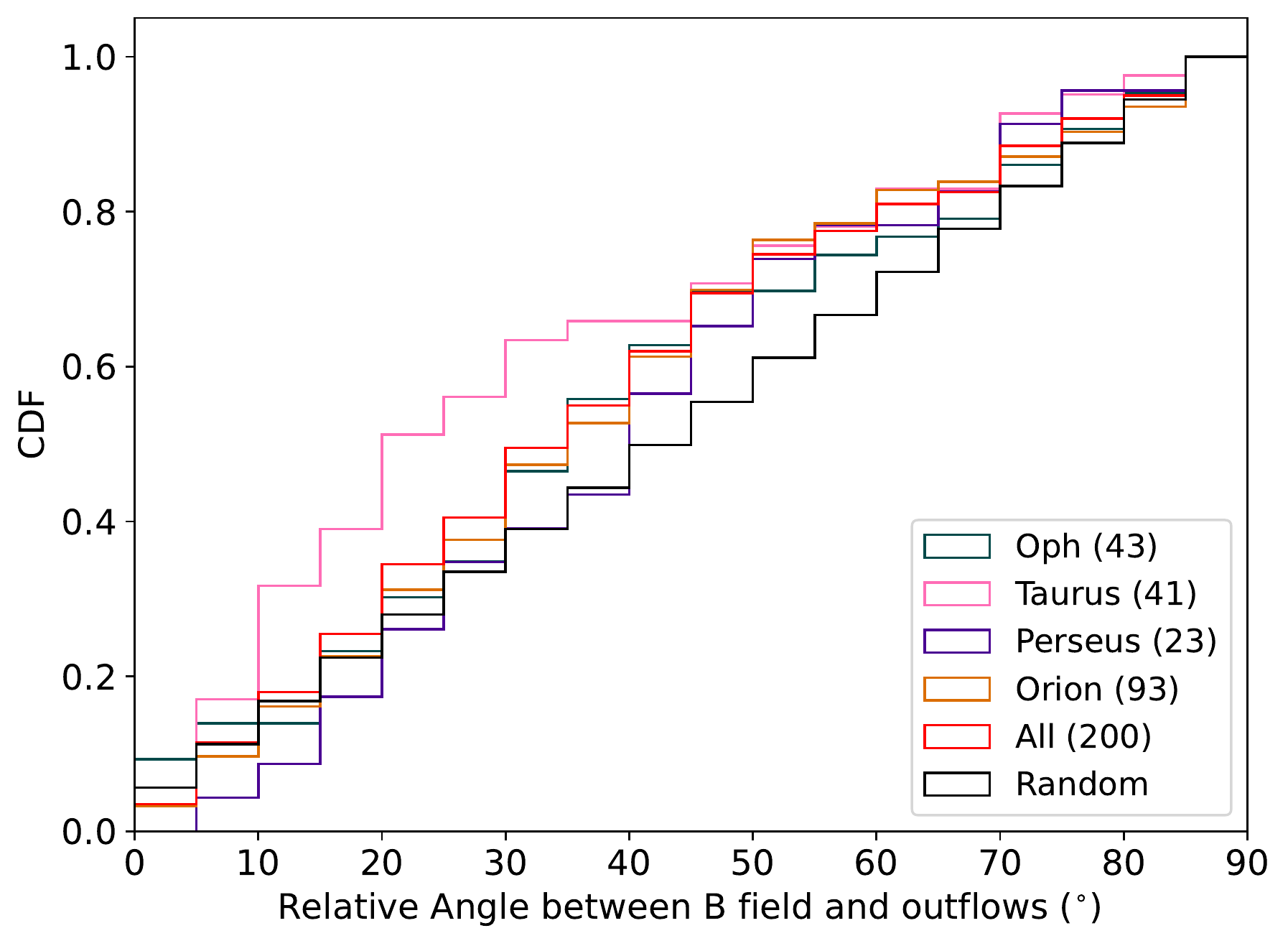}
\includegraphics[width=0.48\linewidth]{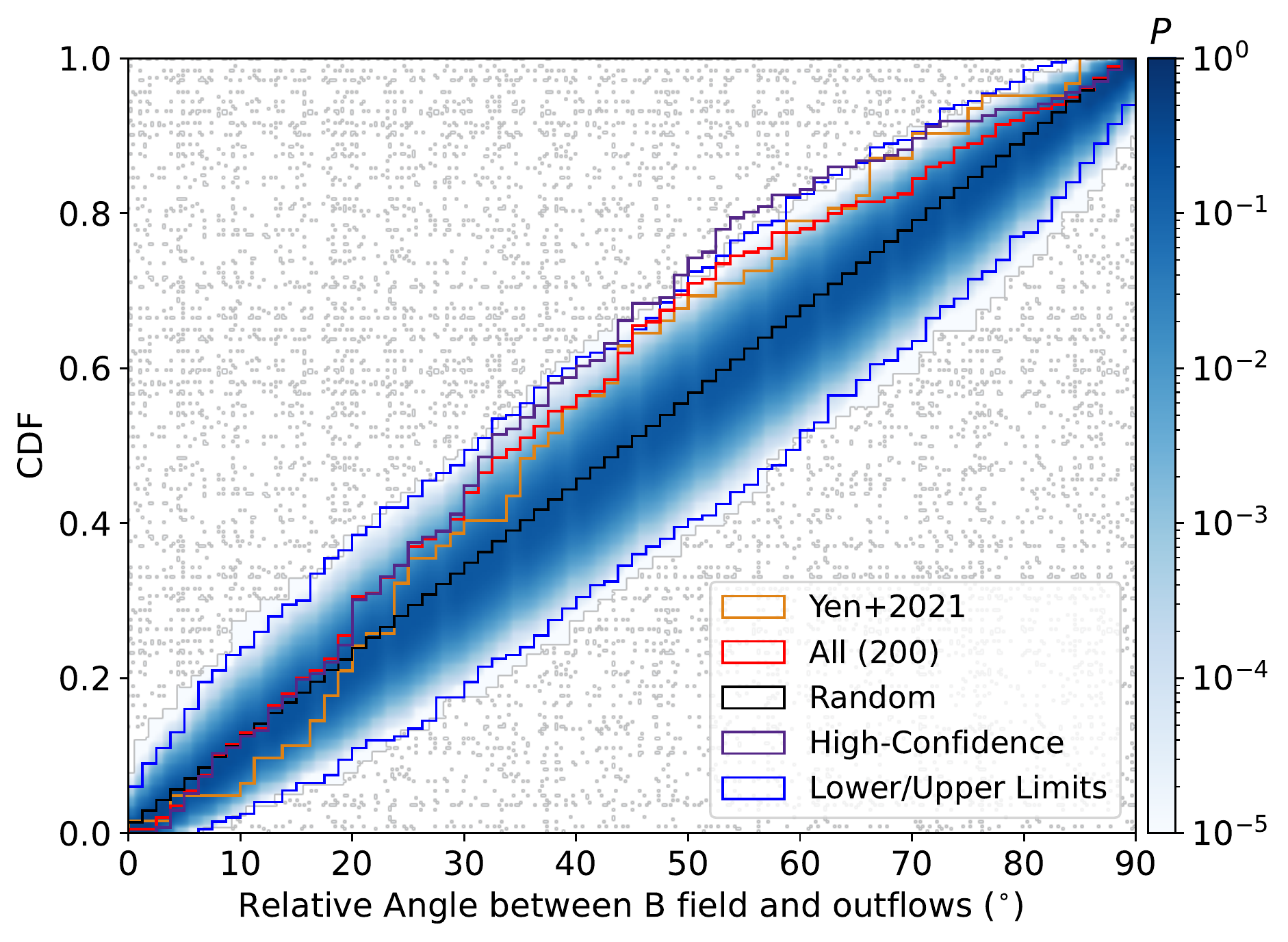}

\caption{{\it (a) Left:} Cumulative distribution function (CDF) of the relative plane-of-sky angle between the magnetic field and outflow orientations, $\theta_{\rm B-out}$, for the 200 protostellar outflow sources in the four cloud regions. The CDF expected from a random distribution of relative angles is also shown. {\it (b) Right:} CDF of $\theta_{\rm B-out}$ for all the outflows overlaid with the resampling test results from a uniform distribution (see text). Two blue lines indicate the upper and lower limits of the CDF when drawing 200 samples from a uniform distribution. The blue color scale illustrates the probability of a CDF drawn from a uniform distribution. The yellow line shows the results from \citet{2021ApJ...907...33Y}. The purple line shows the CDF of $\theta_{\rm B-out}$, for the 136 highest-confidence protostellar outflow sources as discussed in Section~\ref{Outflow Candidates}. }
\label{fig.hist-all-outflow-bfield}
\end{figure*} 

\begin{figure*}[hbt!]
\centering
\includegraphics[width=0.998\linewidth]{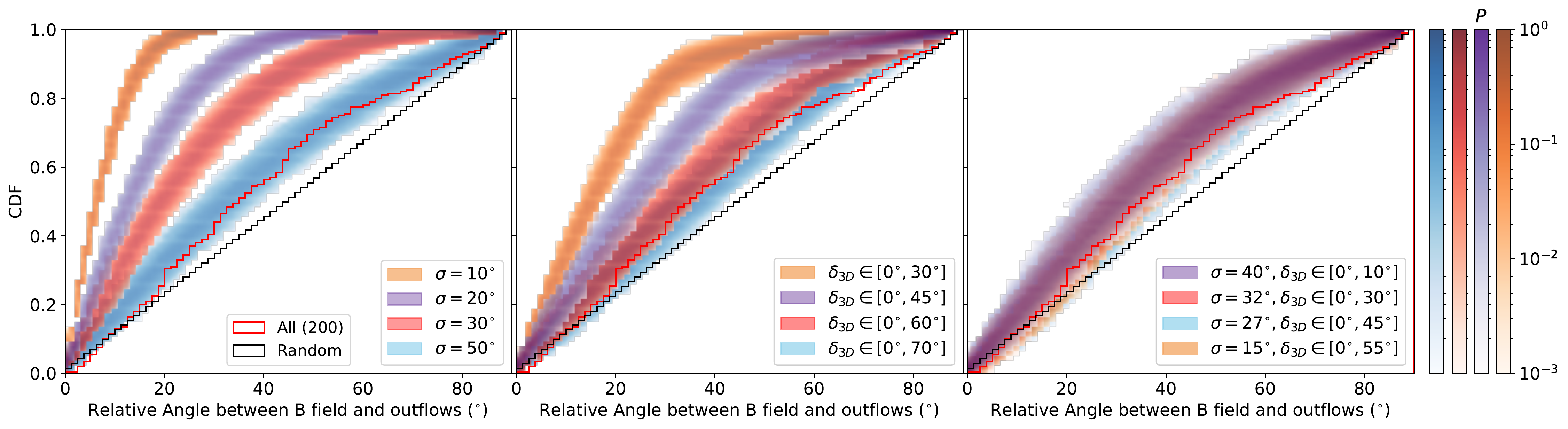}

\caption{Cumulative distribution functions (CDFs) of $\theta_{\rm B-out}$ in mock samples. {\it Left:} $\theta_{\rm B-out}$ distribution of the mock samples with different measurement uncertainties. The uncertainty is the total uncertainty, i.e., that of difference between their orientations on the pane of sky. {\it Middle:} $\theta_{\rm B-out}$ distribution of the mock samples with different 3D orientation angle distributions. {\it Right:} $\theta_{\rm B-out}$ distribution with different measurement uncertainties and different 3D orientation angle distributions. }
\label{fig.hist-all-outflow-bfield-mctest}
\end{figure*} 

In this section, we examine the relative position angle of orientation between magnetic fields and outflows on the plane of sky, $\theta_{\rm B-out}$, for the four regions. We compute the angle between the flux-weighted mean orientation of the magnetic field at the outflow position with the outflow orientation. The angle $\theta_{\rm B-out}$ takes values between $0^{\circ}$ and $90^{\circ}$. The left panel of Figure~\ref{fig.hist-all-outflow-bfield} shows the cumulative distribution function (CDF) of $\theta_{\rm B-out}$ for the four regions. All four regions have a distribution that is above the distribution expected for a random, i.e., uncorrelated, orientation of outflow axis with orientation of magnetic field direction. Note, this random distribution, after accounting for projection effects on the plane of sky, predicts a uniform distribution in $\theta_{\rm B-out}$, so that the CDF rises linearly with $\theta_{\rm B-out}$. 

To evaluate the significance of this result, we conduct a Kolmogorov-Smirnov (K-S) test to examine the difference between the distribution of $\theta_{\rm B-out}$ for all outflows and the uniform distribution that is expected from random relative orientations. The $p$-value of the K-S test is $2.7\times10^{-4}$, which indicates that $\theta_{\rm B-out}$ is not likely to be drawn from a uniform distribution. We then carry out a random sampling test, i.e., we randomly pick 200 numbers from a uniform distribution between 0 and 90, and repeat this process $10^5$ times. The sample size, 200, is given by the number of identified outflows in our study. We show the probability distribution of these samples in the right panel of Figure~\ref{fig.hist-all-outflow-bfield}. The observed $\theta_{\rm B-out}$ distribution is unlikely, with a probability of $\sim 10^{-4}$ that it is drawn from a random distribution. This value is consistent with the $p$-value found in the K-S test.

The right panel of Figure~\ref{fig.hist-all-outflow-bfield} also displays the measured $\theta_{\rm B-out}$ distribution from \citet{2021ApJ...907...33Y}, who selected 62 low-mass protostellar outflows in nearby star-forming regions identified in CO 2-1 and compared their orientations with the magnetic fields measured using JCMT POL-2 data. A two sample K-S test between the \citet{2021ApJ...907...33Y} distribution and our distribution returns a p-value of 0.776, which indicates the two distributions are likely drawn from the same distribution. We also present the $\theta_{\rm B-out}$ distribution of only the highest-confidence outflow candidates, i.e., those have significant coherent high velocity structure in the position-velocity diagram over a range of at least 1 km/s. A two sample K-S test between the \citet{2021ApJ...907...33Y} distribution and the highest-confidence outflow distribution returns a $p$-value of 0.615, which indicates the two distributions are also likely drawn from the same distribution. 
The similarity of the distributions derived from two different data sets and approaches provides further confidence that the distribution of $\theta_{\rm B-out}$ 
is not uniform as would result from random relative orientations. Both distributions peak around $30^{\circ}$, i.e., where the CDF is rising most steeply, and thus have a preference towards alignment rather than misalignment. 

It is worth noting that the magnetic field direction might vary across different physical scales at the same position \citep{2019FrASS...6....3H}. \citet{2021ApJ...907...33Y} adopt the magnetic field traced by JCMT 850 \um, which has a much better spatial resolution (a factor of 10) than that of Planck. Meanwhile, the outflow sample in \citet{2021ApJ...907...33Y} is different from that in this work. Of our 200 outflow candidates, only 22 outflows have high-confident matches with those of \citet{2021ApJ...907...33Y},{which is 35\% of the sample in \citet{2021ApJ...907...33Y} and 11\% of our sample  (see Section~\ref{Compare with Other Work}). On the other hand, the beam size of the Planck data is 4$^{\prime}$.8, which corresponds to a physical scale of 0.2 pc for Ophiuchus and Taurus, 0.4 pc for Perseus and 0.6 pc for Orion. The effective pixel size of the Planck data is 1$^{\prime}$.07, which is four times higher. The typical outflow width is around 0.5~pc \citep{2022ApJ...926...19X}. This indicates that even for the furthest region, Orion, there are at least 16 ($4\times4$) pixels in one outflow. Consequently, the averaged magnetic field direction inside the identified outflow regions will not be significantly affected by the local fluctuation of the magnetic field directions on small scales. Here, we aim to study how large-scale magnetic fields correlate with the launching direction of outflows and leave the examination of smaller scales to future work. }  

Next we explore several possibilities for the origin of the distribution we find in the four regions. We first evaluate how uncertainty in the angle measurement affects the distribution. Here, measurement uncertainties include the ability of the {\it Planck} dust polarization image to measure the true local plane-of-sky $B$-field at the 3D position of source (e.g., it could be affected by other material along line of sight contributing to the overall polarization direction). Measurement uncertainties also include some error in outflow orientations via our method of defining outflow axes, which could also include a contribution from misidentified outflows. To test the impact of measurement uncertainty, we assume the outflow is perfectly aligned with the magnetic field but the measurement of the angles has a Gaussian-distributed uncertainty. We randomly pick 200 numbers from this mock sample, repeating the process $10^3$ times. The left panel of Figure~\ref{fig.hist-all-outflow-bfield-mctest} illustrates the distribution of $\theta_{\rm B-out}$ of the mock samples for four different measurement uncertainties. If either the outflow orientation or the magnetic field direction measurement has about $50^{\circ}$ uncertainty, or alternatively both the outflow orientation and the magnetic field direction measurements have about $35^{\circ}$ uncertainty, the resulting distribution reproduces the observed misalignment distribution. 

We next examine how the distribution of the 3D angle between magnetic field and outflow axis influences the distribution of $\theta_{\rm B-out}$, which is the projected angle. We assume both the outflow orientation and the magnetic field direction are randomly distributed, but we exclude the sample that has a 3D angle over a certain threshold $\delta_{3D}$. This provides a hint about the degree of alignment between the outflow and the magnetic field direction, for example tightly aligned ($0^{\circ}$-$20^{\circ}$) or somewhat aligned ($0^{\circ}$-$45^{\circ}$). These mock samples have a random 3D angle distribution between 0 and $\delta_{3D}$. We then project the 3D angle to 2D from a random viewing angle. The middle panel of Figure~\ref{fig.hist-all-outflow-bfield-mctest} shows the distribution of $\theta_{\rm B-out}$ of the mock samples with different 3D angle distributions. A random distribution with 3D angles between 0$^{\circ}$ and 70$^{\circ}$ also replicates the distribution of $\theta_{\rm B-out}$ we find in this work. 

Finally, we combine both the measurement uncertainty and the 3D angle distribution to investigate how together they can affect the distribution of $\theta_{\rm B-out}$. The right panel of Figure~\ref{fig.hist-all-outflow-bfield-mctest} shows the distribution of $\theta_{\rm B-out}$ with different measurement uncertainties and different 3D angle distributions. Several combinations can lead to the distribution we measure. 
With a moderate angular measurement uncertainty, the magnetic field and outflow could be somewhat aligned. Realistically, we expect at least a 10\% uncertainty both in the field direction and the outflow orientation ($\sigma \sim 15^{\circ}$) \citep{2017ApJ...846...16S,2021ApJ...907...33Y}. 

{ Uncertainty in the outflow orientation comes from several factors. First, the blue- and red-shifted lobes of the outflow may not align with each other. It is not uncommon for two lobes apparently associated with the same source to have different orientations or even appear perpendicular to each other \citep[e.g., some outflows in][]{2010ApJ...715.1170A,2015ApJS..219...20L}. Meanwhile, the morphology for each individual outflow lobe is not necessarily oval or jet-like; distortion may be due to inhomogeneous surroundings, e.g., some outflows in Figure~\ref{fig.hist-all-outflow-bfield-oph}-\ref{fig.hist-all-outflow-bfield-orion}. Outflow directions may also change over time leading to asymmetric morphologies \citep{2016ARA&A..54..491B,2017ApJ...834..201L}. Consequently, it is difficult to fit one high-confident line to the direction of the outflow lobe. To mitigate this we adopt PCA to determine the orientation of the outflow lobes (see Section~\ref{Outflow Candidates}). For some outflow lobes, the variance of the two orthogonal components are similar, which indicates a large uncertainty in the outflow orientation estimation. 
Likewise, the magnetic field direction inferred from the dust polarization contains several systematic uncertainties, including smoothing, contamination by CMB polarization, and leakage correction as discussed in \citet{2015A&A...576A.106P}. Consequently, it is reasonable to expect $\sigma \sim 15^{\circ}$ uncertainty at minimum in both the field direction and the outflow orientation.} In this case the curve with  $\delta_{\rm 3D} \in [0^{\circ},55 ^{\circ}]$ provides a good match to the observed distribution. Likely the total measurement uncertainty is larger. 

Consequently, we conclude that large scale magnetic fields play some role in regulating the launching direction of outflows and governing the eventual extent of the CO-traced entrained gas. {However, the magnitude of their importance is somewhat degenerate with the accuracy at which the outflow and magnetic field directions can be determined.}



\subsection{Outflow Pair Relative Orientations versus Separation}
\label{Outflow Pair Misalignment versus Separation}

\begin{figure*}[hbt!]
\centering
\includegraphics[width=0.32\linewidth]{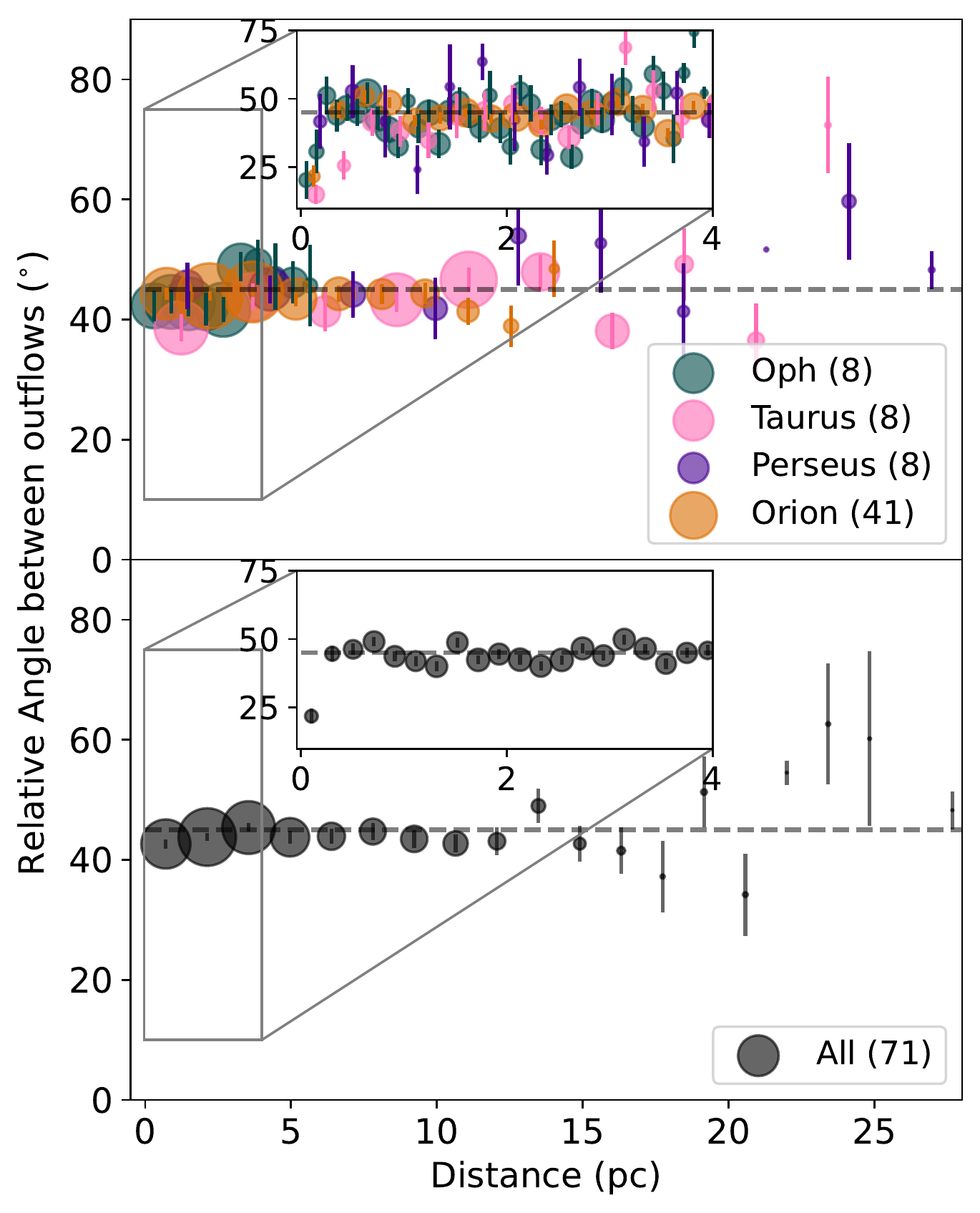}
\includegraphics[width=0.32\linewidth]{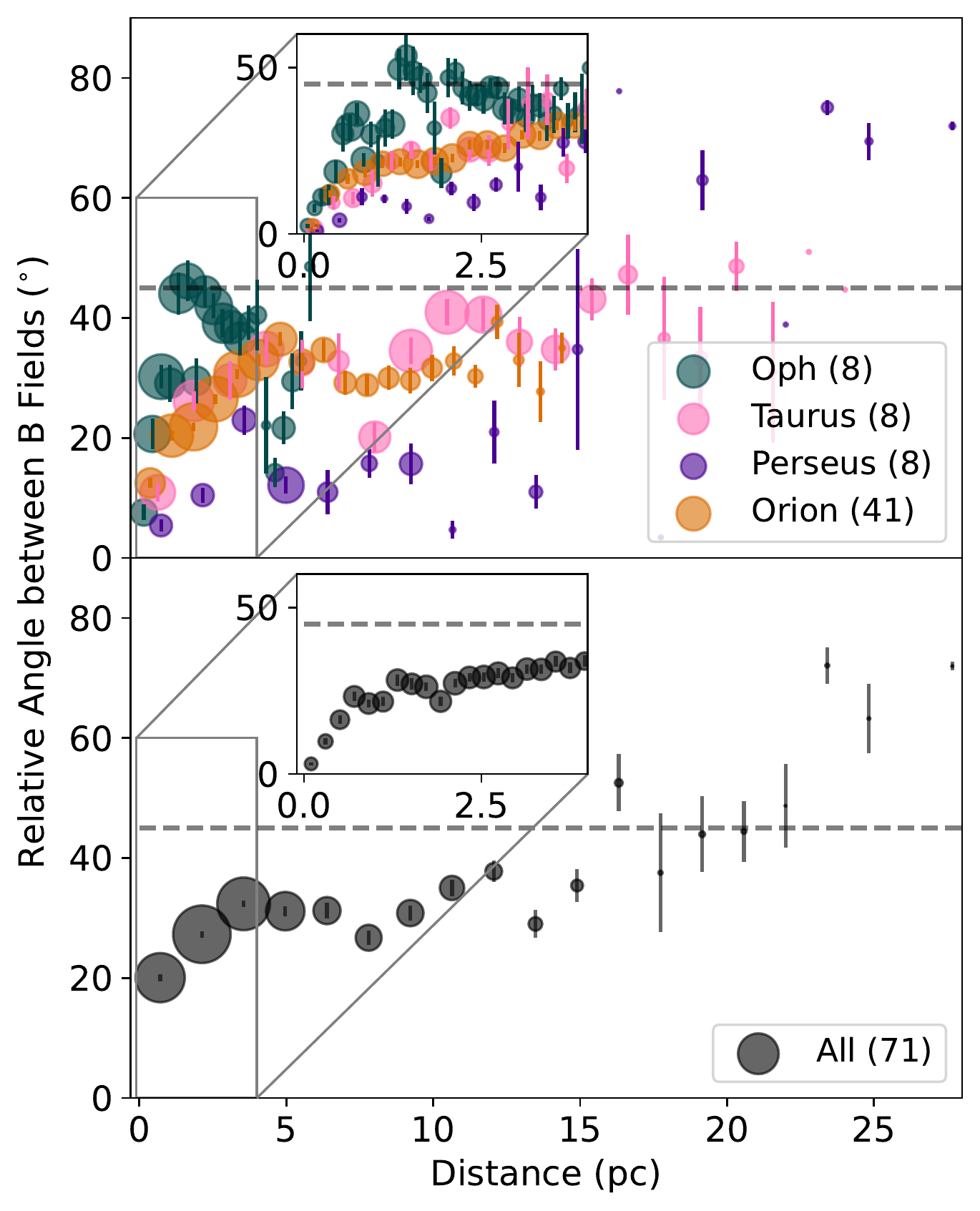}
\includegraphics[width=0.32\linewidth]{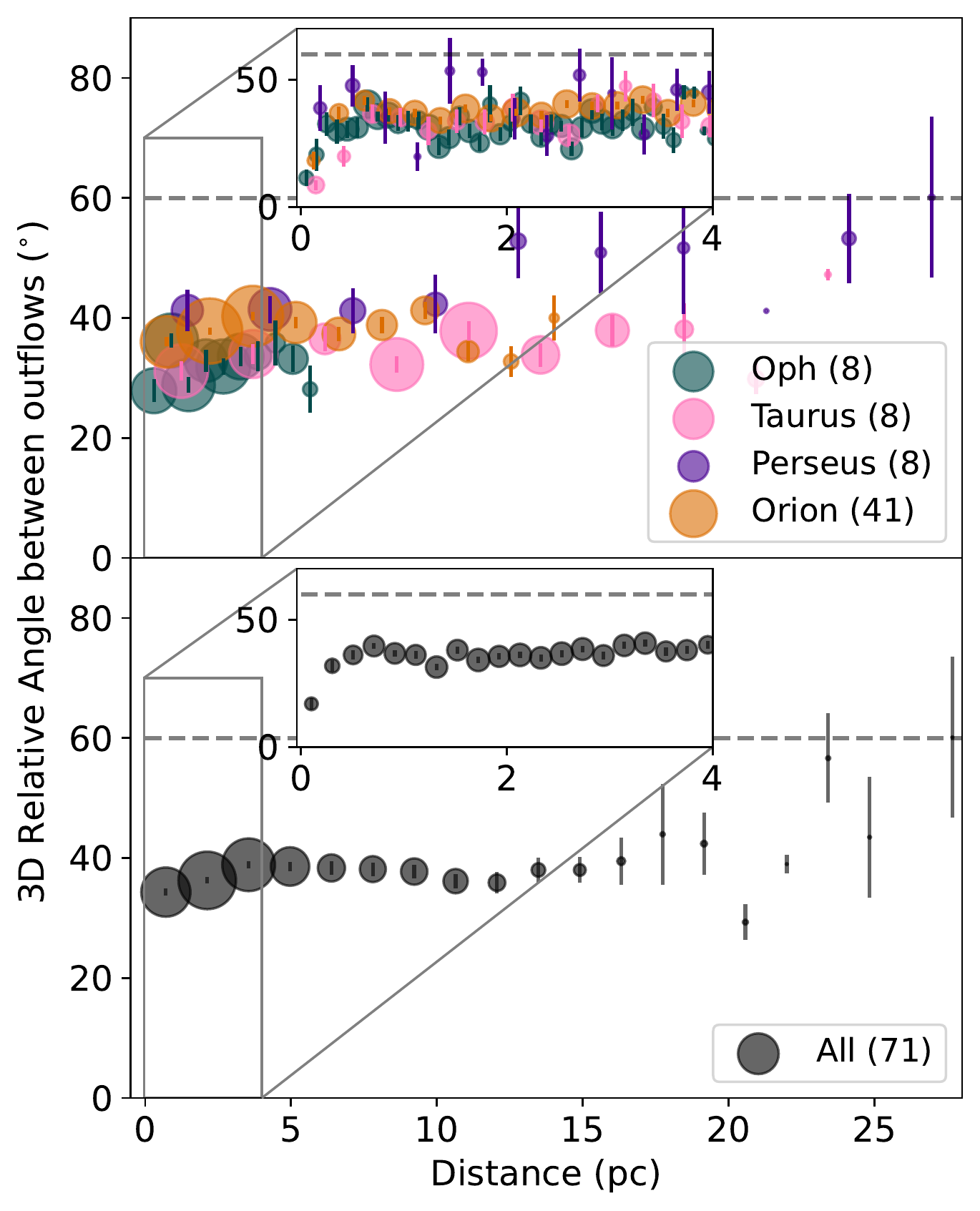}

\caption{{\it Left:}  Plane-of-sky relative orientation angles of outflow pairs as a function of their projected separation distances. {\it Middle:} Plane-of-sky relative orientation angles of the local magnetic field directions at the positions of outflow pairs as a function of their projected separation distances. {\it Right:} 3D relative orientation angles of outflows pairs as a function of their projected separation distances. The data have been averaged in various separation distance bins, with the size of the circle indicating the number of samples in each distance bin (see legend). The gray dashed line illustrates $45^{\circ}$ (left and middle panels) and $60^{\circ}$ (right panel), which is expected from random relative orientations. The inset shows a zoom to the smallest scales from 0 to 4~pc.
}
\label{fig.hist-all-outflow-self-distance-all}
\end{figure*}



In this section, we investigate how the orientations of outflow pairs are correlated with the projected separation distance of the sources. We calculate the plane-of-sky orientation angle difference between all outflows with all other outflows. For each bin of projected separation distance, we calculate the mean relative orientation angle and the standard deviation. The left panel of Figure~\ref{fig.hist-all-outflow-self-distance-all} illustrates how the relative orientation angle between outflow pairs changes with separation for the four regions. At small separations, $\lesssim 0.5\:$pc, the average relative orientation angles are $<45^\circ$, indicating a modest degree of alignment {but with significant dispersion about the perfectly aligned orientation}. At larger separations, the average relative orientation angle is close to  $45^{\circ}$, implying that the relative orientation angles have become decorrelated on these scales.

Prior work examining the alignment of  protostellar outflow pairs with projected separations of $\sim 1,000-9,000$ AU  found that the angle distribution is more consistent with an anti-aligned distribution \citep{2016ApJ...820L...2L}. Similarly, the angle difference of binaries measured in star formation simulations is statistically consistent with a random distribution \citep{2016ApJ...827L..11O,2019ApJ...887..232L}. Synthetic CO observations demonstrate that the apparent anti-alignment may be produced by a selection affect, whereby aligned outflows are more difficult to identify and separate  \citep{2016ApJ...827L..11O,2019ApJ...887..232L}. Likewise, this effect would make the distribution of relative angles {\it more} aligned at close separations than we report here. We note the size scale investigated by these studies is slightly below our present resolution. On these scales, it is also likely that the alignment is influenced by dynamical interactions, which erase the initial outflow orientations  \citep[e.g.,][]{2017ApJ...834..201L}. Since the closest pairs in our study have separations larger than the typical core size, they are not wide-separation companions. This means our results may be less influenced by dynamical evolution and instead better reflect the initial configuration.

As shown in Figure~\ref{fig.hist-all-outflow-bfield-oph}-\ref{fig.hist-all-outflow-bfield-orion}, magnetic fields are more ordered on smaller scales ($<5$ pc) and tend to have a similar orientation, such that the relative position angle between the magnetic field at two positions is smaller for closer distances. The middle panel of Figure~\ref{fig.hist-all-outflow-self-distance-all} shows how the relative angle of the magnetic field direction at the corresponding outflow positions changes with separation for the four regions. At small separations, $\lesssim 2\:$pc for Ophiuchus and $\lesssim 5\:$pc for the other three regions, the relative orientation angles increase with the separation, indicating self-correlated magnetic field directions. However, this trend vanishes at larger separations, implying that the relative angles between magnetic field directions become decorrelated on these scales. It is worth noting that the beam size of the Planck data corresponds to a physical scale of 0.2 pc for Ophiuchus and Taurus, 0.4 pc for Perseus and 0.6 pc for Orion. 

{We note that for Perseus the relative angle between polarization vectors, noted as $\theta_{\rm B-B}$, behaves differently than the others. It first increases within 4 pc, but decreases and maintains low values between 5 to 15 pc, then significantly increases above 45 $\circ$ after 15 pc. This unusual behavior might be caused by feedback or due to some particular magnetic field morphology on these scales. However, Orion, which is the most active star-forming region, does not show a similar trend. As we discussed in Section~\ref{Outflow Orientation Distributions}, the plane-of-sky magnetic field strength in Perseus is likely the weakest among the four regions, while Orion seems to have a strong magnetic field \citep[e.g.,][]{2021ApJ...913...85H}. 
When the magnetic field is strong, $\theta_{\rm B-B}$ is likely small across all separations. Due to turbulence perturbations, $\theta_{\rm B-B}$ likely increases with distance and becomes ``random" ($\sim 45 ^{\circ}$) at larger scales. Perseus exhibits an irregular change in $\theta_{\rm B-B}$ with distance. At smaller distances, $\theta_{\rm B-B}$ of Perseus is small, as expected, possibly indicating low turbulence or stronger magnetic fields. However, at larger distances, its unusual behavior can not be explained by strong magnetic fields. Consequently, whether Perseus has weaker magnetic field relative to the other regions is still not clear.  

}


Comparing the left and middle panels of Figure~\ref{fig.hist-all-outflow-self-distance-all} indicates that the angle differences between outflow pairs are less well correlated at small scales than those of the magnetic field directions at the same locations. This may be because outflow orientations change on shorter timescales due to dynamics or due to small scale turbulence, which is not reflected in the mean magnetic field direction.
We conclude that these results are consistent with the magnetic field influencing the direction of outflow launching and/or propagation (as found in Section \ref{Outflow Orientation versus Magnetic Field Directions}), but with a high dispersion around 
aligned orientations. This large dispersion means that the outflow to outflow alignment signal decorrelates more rapidly with distance than the $B$-field to $B$-field alignment signal and is soon indistinguishable from a random distribution.



Next, we adopt the inclination angles with respect to the line of sight inferred with a \CASItD\ in Section ~\ref{sec-CASItD}. We calculate the 3D launching direction of each outflow by combining both the position angle and the inclination angle. We calculate the 3D outflow angle difference for all outflow pairs in the clouds and bin by separation. The right panel of Figure~\ref{fig.hist-all-outflow-self-distance-all} illustrates how the 3D relative angle between outflow pairs changes with separation for the four regions. We find this angle shows a similar behavior to the projected relative angle versus separation relation shown in the left panel of Figure~\ref{fig.hist-all-outflow-self-distance-all}, i.e., there is enhanced correlation within $\sim 0.5\:$pc, but then a relatively constant distribution of the average relative angle with projected separation distance. We also note that this constant value of average angle is significantly smaller than the value expected for a random distribution in 3D space, which is $60^\circ$. We attribute this to a detection bias in the inclination angle distribution that selects against certain angles that are either close to the plane of the sky or along the line of sight (see Figure~\ref{fig.cdf-inclination-four-region}). The existence of such a bias implies that results for the correlation of 3D orientations need to be treated with caution. 




{


\subsection{Comparisons with Other Work}
\label{Compare with Other Work}
 
 There have been several prior studies investigating how outflows are orientated with respect to filamentary structures. Assuming that the magnetic field direction correlates with the filament morphology, the filament orientation provides some ancillary insight into the local field behavior.
 \citet{2016A&A...586A.135P} showed that at low column densities ($N({\rm H})\sim 10^{20}\,{\rm cm}^{-2}$), 
 structures are preferentially aligned with the magnetic field inferred from the polarization angle. However, at high column densities ($N({\rm H})\sim 10^{22}\,{\rm cm}^{-2}$), e.g., in molecular clouds, gas structures are preferentially perpendicular with the magnetic field. Therefore, in the regime we study it is likely that the magnetic fields are perpendicular to the filamentary structures. This would suggest that the outflows we identify are more likely to be oriented perpendicular to the local filament direction.
 
 Indeed, \citet{2019ApJ...874..104K} found that most outflows are preferentially perpendicular to the filaments in IRDC G28.37+0.07, rejecting the random distribution at high confidence, which is similar to our work. \citet{2020ApJ...896...11F} found a similar trend for the highest-confidence outflows in Orion, which are associated with driving sources and correlated with \h2\ 2.122 \um\ outflow emission. They exhibit a moderately perpendicular outflow-filament alignment. However, when considering all outflow samples in Orion, \citet{2020ApJ...896...11F} found random outflow-filament alignment, which is similar to the findings in Perseus \citep{2017ApJ...846...16S}. One possible explanation is that stronger feedback or turbulence leads to more random alignment, while stronger magnetic fields tend to produce more alignment. Since Perseus may have a relatively weak magnetic field as discussed in Section~\ref{sec-CASItD} and \ref{Outflow Pair Misalignment versus Separation}, this might explain why \citet{2017ApJ...846...16S} found a random outflow-filament alignment in Perseus.  
 
 In this section, we also report a one-to-one comparison with the outflow position angles and the magnetic field directions reported in \citet{2021ApJ...907...33Y}. \citet{2021ApJ...907...33Y} adopted 62 outflows previously identified in either \co\ (2-1) or \co\ (3-2) observations by JCMT and SMA and compared the outflow orientation with the mean magnetic field direction measured from JCMT POL-2 data. To identify sources contained in both catalogs we conduct a close companion search using the outflow coordinates. We define matches as those with separations within 5$^{\prime}$. This identifies 22 outflow matches. We find that 54\% of the outflow position angle measurements between the two samples have orientations within 20$^{\circ}$. However, 18\% of the outflow position angle measurements have discrepancies greater than 70$^{\circ}$. The magnetic field direction measurements between the two samples exhibit similar behavior, with 45\% of magnetic field orientations within 20$^{\circ}$ and 18\%  with over 70$^{\circ}$ discrepancy. We expect that most of this discrepancy arises from the different datasets used to identify the outflows: the morpologies of the outflows are simply different and the magnetic field directions are measured on different scales.
 While we expect that the higher resolution data of \citet{2021ApJ...907...33Y} allows a more exact determination of the outflow position angle, we nonetheless find a statistically similar distribution and reach the same conclusion.


}

\section{Conclusions}
\label{Conclusions}

We have used supervised machine learning to identify a large sample of outflows that have orientations defined by their blue and redshifted velocity components in several nearby molecular clouds (Ophiuchus, Taurus, Perseus and Orion). We use the sample to study the correlation between outflow orientation and magnetic field direction. We have also developed a new convolutional neural network model to predict the inclination angle of outflows with respect to the line-of-sight. Our main findings are as follows:

\begin{enumerate}

\item  The plane-of-sky orientations of outflows show a preference towards alignment with the plane-of-sky magnetic field as measured by $Planck$ observations of dust polarization. The significance of the alignment signal is high, i.e., there is only a small probability, $\sim 10^{-4}$, that the distribution is consistent with random orientations of the outflows with respect to the local magnetic fields. However, the distribution of relative orientation angles peaks around $30^{\circ}$. 
The distribution can be explained as some combination of measurement uncertainties (i.e., in estimating the true plane-of-sky $B$-field at the protostar's position and its true plane-of-sky outflow orientation from our method of defining outflow axes from blue and redshifted CO emission) and intrinsic deviation from perfect alignment.
Our observed distribution is consistent with the previous study of \citet{2021ApJ...907...33Y}, which analyzed a different, smaller sample of outflows and utilized a different magnetic field survey that probed fields on smaller, more local scales. The physical implications of this result are that magnetic fields have a dynamical influence on the direction of outflow launching and/or propagation. However, there may be other physical processes, e.g., turbulence or strong stellar dynamical interactions, that can also significantly affect the outflow orientations, i.e., by inducing significant deviations from perfect alignment of outflows with their local $B$-fields.

\item The distribution of plane-of-sky relative orientations between outflow pairs shows an alignment signal only on small scales, i.e., at projected separation distances of $\lesssim 0.5\:$pc, even though the $B$-field to $B$-field orientation shows correlation out to larger scales. This rapid decorrelation in outflow relative orientations with distance is further evidence for the high degree of scatter in the orientation of individual outflows with respect to the local $B$-fields.


\item Our \CASItD\ model is able to predict the inclination angle of outflows with respect to the line of sight with an uncertainty of $\sim 10 ^{\circ}$. The average inclination angle of outflows in Perseus ($\sim 54^{\circ}$) is larger than that of the other three regions ($\sim 36^{\circ}-39^{\circ}$). However, it is likely that the method of outflow detection leads to biases in the inclination angles that are selected, i.e., disfavoring plane-of-sky orientations and potentially near pole-on orientations.
\end{enumerate}

{Our work motivates further study to examine the correlation between the outflow and magnetic fields with higher-resolution observations. This will place firmer constraints on the uncertainty of the outflow directions as well as the magnetic field directions. Meanwhile, in simulations, investigating different magnetic field strengths will provide more insight into the role of magnetic fields in setting the outflow direction. In addition, comparing different treatments for the outflow launching direction in simulations, e.g., aligned with the angular momentum of the sink particle \citep{2011ApJ...740..107C,2021MNRAS.506.2199G} or simply aligned with local magnetic fields \citep{2010ApJ...709...27W}, is a crucial step to understand the impact of feedback on the natal cloud. 

}

D.X., S.S.R.O., and R.A.G. acknowledge support by NSF grant AST-1812747. D.X. acknowledges support from the Virginia Initiative on Cosmic Origins (VICO). J.C.T. acknowledges support from NSF grants AST-1910675 and AST-2009674 and ERC project MSTAR. The Texas Advanced Computing Center (TACC) at the University of Texas at Austin provided HPC resources that have contributed to the research results reported within this paper. The authors acknowledge Research Computing at The University of Virginia for providing computational resources and technical support that have contributed to the results reported within this publication.

\appendix

\section{\CASItD\ Performance}
\label{CASI Performance}

\begin{figure*}[hbt!]
\centering
\includegraphics[width=0.68\linewidth]{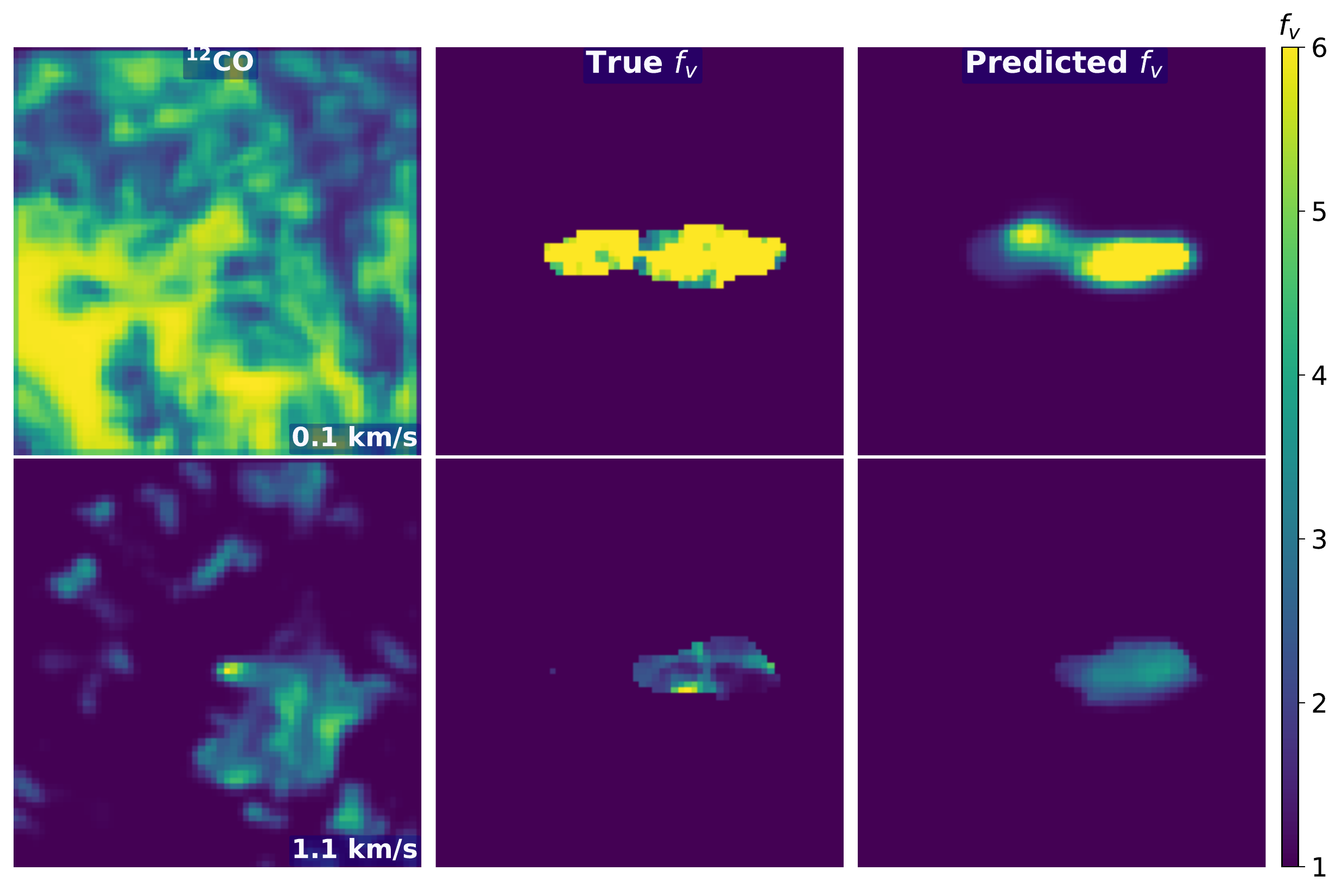}
\caption{\co\ emission of a synthetic outflow at two velocity channels (left column), the corresponding correction factor (middle column), and \CASItD\ prediction (right column). }
\label{fig.outflow-simulation-outflow-example-casi}
\end{figure*}

In this section, we evaluate the performance of \CASItD\ on both synthetic data and \co\ (1-0) observations. As discussed in Section~\ref{sec-CASItD}, we train a new \CASItD\ model to predict $f_{v} =\frac{P_{3D}}{P_{LOS}}$, which is the correction factor between the actual 3D momentum to the LOS momentum in each pixel. Figure~\ref{fig.outflow-simulation-outflow-example-casi} shows an example of a synthetic outflow at two velocity channels and the corresponding correction factor. We also show the prediction by \CASItD\ in Figure~\ref{fig.outflow-simulation-outflow-example-casi}. The correction factor is large at low velocity channels and decreases when the LOS velocity increases. The \CASItD\ model is able to capture this trend and predict the correction factor accurately.

\begin{figure}[hbt!]
\centering
\includegraphics[width=0.48\linewidth]{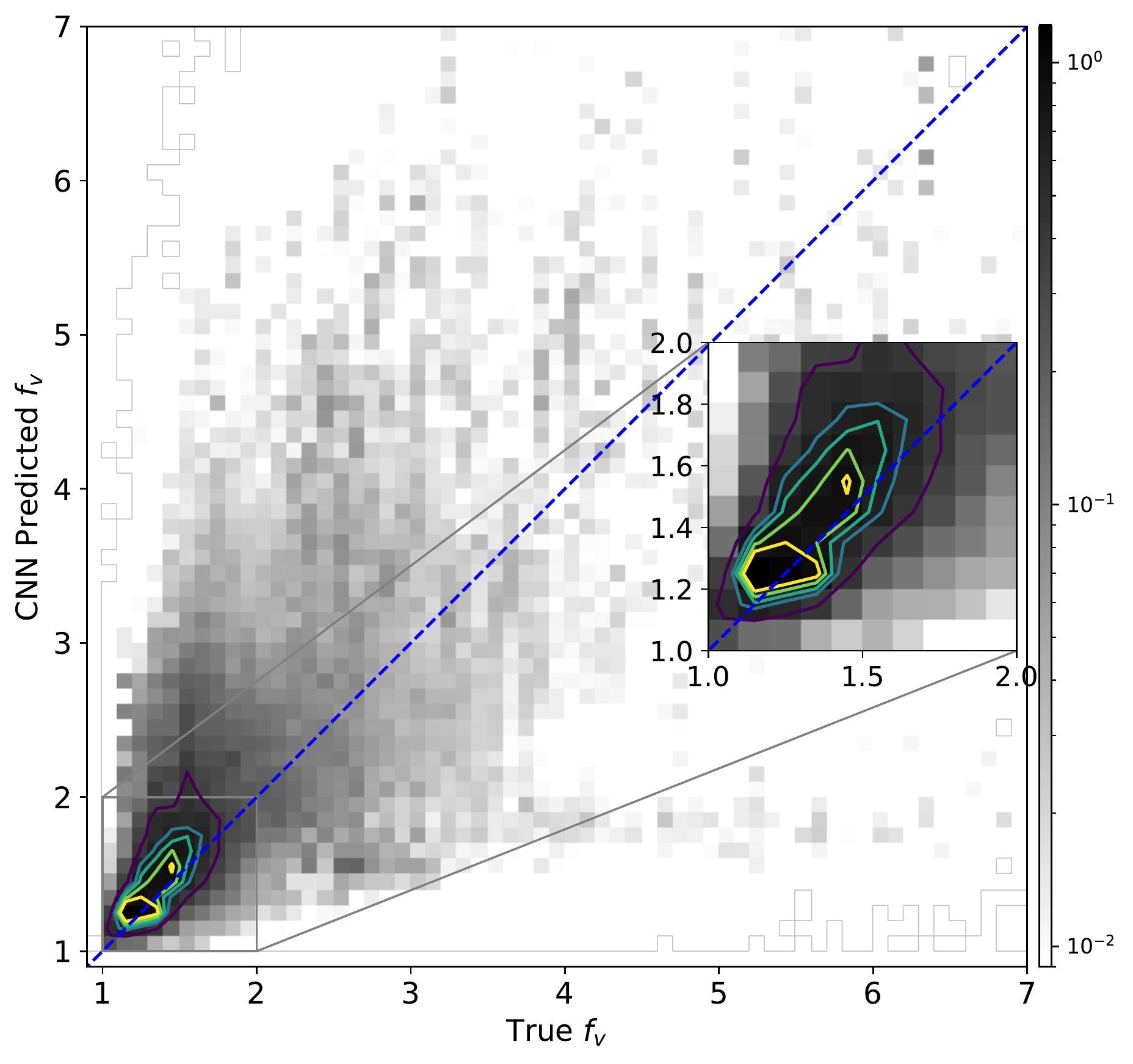}
\caption{Scatter plot between true $f_{v}$ and \CASItD\ predicted  $f_{v}$. The color indicates the normalized occurrence of the data point. }
\label{fig.pixel-stat-velfrac}
\end{figure}

Figure~\ref{fig.pixel-stat-velfrac} demonstrates the performance of \CASItD\ for predicting the correction factor for each pixel in the spectral cube. Although there is some scatter between the true $f_{v}$ and the \CASItD\ predicted $f_{v}$, the darkest region, where most data points are situated, is located on the one-to-one line. This demonstrates that \CASItD\ correctly predicts the correction factor on average. We derive the inclination angle of the outflow in each data cube by taking the average of the individual pixel predictions for $f_{v}$  (see Figure~\ref{fig.test-inclination-errorbar}). \CASItD is able to predict the inclination angle of outflows within an uncertainty of $\sim 10 ^{\circ}$.

\bibliographystyle{aasjournal}
\bibliography{references}

\end{CJK*}

\end{document}